\documentclass[final,5p,times,twocolumn]{elsarticle}

\usepackage{graphicx}
\usepackage{amssymb}
\usepackage{amsmath}
\usepackage{physics}
\usepackage{lineno}
\usepackage{rotating}
\usepackage{xcolor}
\usepackage{mhchem}
\usepackage{siunitx}
\usepackage{orcidlink}
\usepackage{subcaption}

\journal{NIM A}

\begin{document}

\begin{frontmatter}

\title{The BigBite Calorimeter for the Super Bigbite Spectrometer Program at Jefferson Lab} 



\author[UCONN,LBNL]{P.~Datta\corref{cor}\orcidlink{0000-0001-5954-8204}}
\cortext[cor]{Corresponding author}
\ead{pdatta@lbl.gov}
\author[WM]{K.T.~Evans\orcidlink{0000-0001-5498-8008}}
\author[UMASS]{J.~Bane\orcidlink{0000-0003-2199-9733}}
\author[MSU]{H.~Bhatt\orcidlink{0000-0003-0087-5387}}
\author[MSU]{B.~Devkota\orcidlink{0000-0002-2859-5961}}
\author[UCONN,WM]{E.~Fuchey\orcidlink{0000-0003-0100-6052}}
\author[JLAB]{T.~Hague\orcidlink{0000-0003-1288-4045}}
\author[JLAB]{D.W.~Higinbotham\orcidlink{0000-0003-2758-6526}}
\author[JLAB]{A.~Hoebel}
\author[JLAB]{M.K.~Jones\orcidlink{0000-0002-7089-6311}}
\author[MSU]{A.~Karki\orcidlink{0000-0001-7650-2646}}
\author[ITEP,NWU]{M.~Kubantsev}
\author[UNH]{S.~Li\orcidlink{0000-0003-1252-5392}}
\author[UVA]{M.~Nycz\orcidlink{0000-0002-3570-9103}}
\author[KHAR]{R.~Pomatsalyuk\orcidlink{0000-0002-6299-7446}}
\author[UCONN]{A.J.R.~Puckett\orcidlink{0000-0002-3639-7463}}
\author[BUD]{I.~Rachek\orcidlink{0000-0001-7819-7534}}
\author[ARG]{S.~Riordan}
\author[JLAB]{B.~Sawatzky\orcidlink{0000-0002-5637-0348}}
\author[UCONN]{S.A.~Seeds\orcidlink{0000-0001-6469-6607}}
\author[YER]{A.~Shahinyan}
\author[BUD]{Y.~Shestakov}
\author[JLAB]{A.S.~Tadepalli\orcidlink{0000-0002-5312-8943}}
\author[ITEP]{\textdagger V.~Verebryusov}
\author[YER]{H.~Voskanyan\orcidlink{0000-0001-9515-3568}}
\author[JLAB]{B.~Wojtsekhowski}
\author[CNU, JLAB]{A.~Yoon}

\address[UCONN]{\mbox{University of Connecticut, Storrs, CT 06269, USA}}
\address[LBNL]{\mbox{Lawrence Berkeley National Laboratory, Berkeley, CA 94720, USA}}
\address[WM]{\mbox{William \& Mary, Williamsburg, VA 23187, USA}}
\address[UMASS]{\mbox {University of Massachusets, Amherst, MA, USA}}
\address[MSU]{\mbox{Mississippi State University, Mississippi State, MS 39762, USA }}
\address[JLAB]{\mbox{Thomas Jefferson National Accelerator Facility,~
Newport News, VA 23606, USA}}
\address[ITEP]{\mbox{Institute of Theoretical and Experimental Physics, Moscow, Russia }}
\address[NWU]{\mbox{Northwestern University, Chicago, IL, USA}}
\address[UNH]{\mbox{University of New Hampshire, Durham, NH, USA}}
\address[UVA]{\mbox {University of Virginia, Charlottesville, VA, USA}}
\address[KHAR]{\mbox {Kharkov Institute of Physics and Technology, Kharkov, Ukraine}}
\address[BUD]{\mbox{Budker Institute of Nuclear Physics, Novosibirsk, Russia}}
\address[ARG]{\mbox{Argonne National Laboratory, Lemont, IL, USA}}
\address[YER]{\mbox{Yerevan Physics Institute, Yerevan, Armenia }}
\address[CNU]{\mbox{Christopher Newport University, Newport News, VA 23606, USA}}


\begin{abstract}
    We report features of the design, construction, installation, and performance of the BigBite Calorimeter (BBCal), a lead-glass electromagnetic calorimeter constructed as part of the BigBite Spectrometer (BBS), which served as the electron arm for the Super Bigbite Spectrometer (SBS) program of high-precision neutron electromagnetic form factor measurements in Hall A at Jefferson Lab. As a total-absorption calorimeter, BBCal provided the primary electron trigger for BBS, detecting (quasi-) elastically scattered electrons in the 1-4~GeV energy range with an energy resolution of approximately $6.2\%$, position resolution of 1.2~cm, and timing resolution of 0.5~ns. 
\end{abstract}

\begin{keyword}
    Calorimeter \sep Shower \sep Preshower \sep BBCal \sep Nucleon Form Factors
\end{keyword}

\end{frontmatter}

\section{Introduction}
\label{sec:introduction}




The first version of the BigBite calorimeter (BBCal) was constructed during the 6 GeV era of Jefferson Lab experiments aimed at exploring nucleon structure in exclusive and semi-inclusive processes in the intermediate $Q^2$ range ~\cite{Riordan:2010id,JeffersonLabHallA:2011ayy}. Due to the drop in cross section, the next generation of experiments that wished to extend these studies to larger $Q^2$ values required detectors which could handle higher luminosities, requiring higher data acquisition and trigger rate capabilities, and better, radiation-hard calorimeter modules. The advent of high-luminosity polarized targets, fast tracking detectors, and high-intensity electron beams opened doors to a new series of high-precision measurements. 


The 12 GeV upgrade of the Continuous Electron Beam Accelerator Facility (CEBAF) \cite{PhysRevAccelBeams.27.084802} has enabled a new generation of high-precision measurements at Jefferson Lab (JLab). Taking advantage of the upgrade to the beam energy and detector capabilities, the Super Bigbite Spectrometer (SBS) collaboration conducted a series of experiments in JLab's experimental Hall A between fall 2021 and summer 2025 to determine nucleon electromagnetic form factors with unprecedented precision and resolution. The objectives of the experiments that used the calorimeter described in this paper are summarized below:


\begin{itemize}
    \item \textbf{E12-09-019 (SBS-GMn, Oct 2021–Feb 2022)} — First SBS experiment, extending high-precision measurements of the neutron magnetic form factor $G_M^n$ over $Q^{2}=3$-13.6~(GeV/$c$)$^{2}$ using the ratio method \cite{bg01}.

    \item \textbf{E12-20-010 (SBS-nTPE, Jan 2022–Feb 2022)} — Ran in series with SBS-GMn to perform the first Rosenbluth separation of the neutron form factors at high $Q^{2}$~($\sim$4.5~(GeV/$c$)$^{2}$) \cite{bg02}.

    \item \textbf{E12-09-016 (GEn-II, Oct 2022–Oct 2023)} — Extended precision measurements of the neutron electric form factor $G_E^n$ over $Q^{2}=3$–10~(GeV/$c$)$^{2}$ via beam–target double spin asymmetry technique utilizing a state-of-the-art polarized ${}^3$He target \cite{ca09,Helium3_target2015}.

    \item \textbf{E12-17-004 (GEn-RP, Apr–May 2024)} — Performed a $G_E^n$ measurement via the recoil polarization technique at high $Q^{2}$~($\sim$4.5~(GeV/$c$)$^{2}$). Comparison with E12-20-010 and E12-09-016 results will provide critical tests of the one-photon exchange (OPE) approximation in elastic $en$ scattering \cite{bg03}.
    
     \item \textbf{E12-20-008 ($K_{LL}$, May 2024)} — Performed a measurement of polarization transfer in charged pion photoproduction in the wide angle regime. Data from this experiment will provide a critical test of the applicability of the handbag mechanism \cite{bg05}.

\end{itemize}


All experiments listed above employed a two-arm coincidence setup. The hadron arm consisted of the Super Bigbite Spectrometer (SBS), while the BigBite Spectrometer (BBS) was used as the electron arm. In the SBS, scattered nucleons were momentum-analyzed by the SBS dipole magnet and detected in a highly segmented hadron calorimeter (HCal) with comparable and high efficiencies for protons and neutrons. The electron arm used the BigBite dipole magnet to bend scattered charged particles into or out of the acceptance depending on their charge and momentum. Precise tracking in the BBS was achieved with five layers of Gas Electron Multiplier (GEM) detectors, and energy measurements were made in BBCal. Additional BBS subsystems included the Gas Ring Imaging Cherenkov (GRINCH) detector for particle identification and a timing hodoscope (TH) for time-of-flight measurements.

\begin{figure}[h]
    \centering
        \includegraphics[width= 0.95\linewidth]{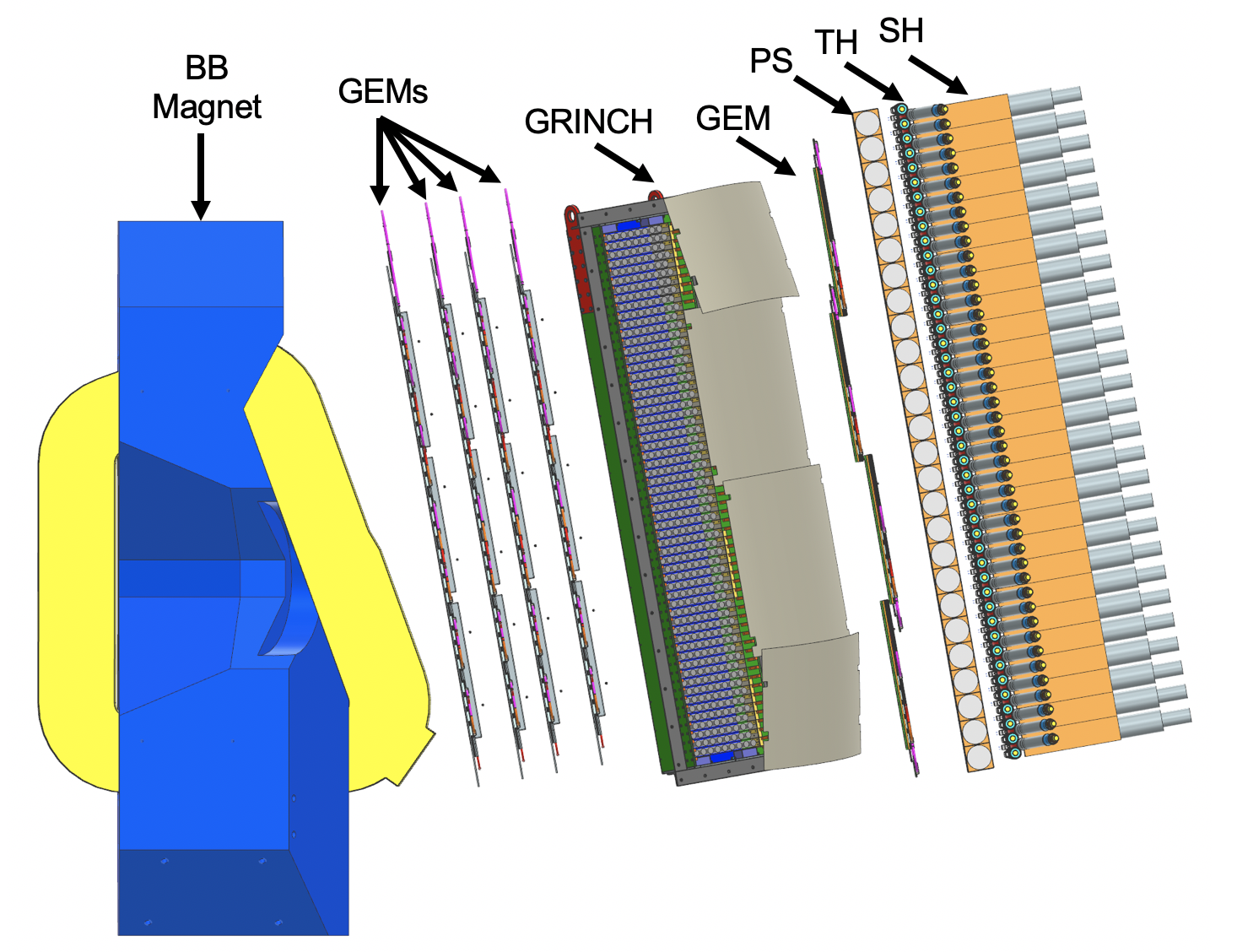}
    \caption{Side view CAD drawing of the BBS detector stack installed in experimental Hall A. Scattered electrons travel through the detector stack from left to right in the drawing. Here, BB Magnet is the BigBite magnet, GEMs are the Gas Electron Multiplier detectors, GRINCH is the Gas Ring Imaging CHerenkov detector, PS is preshower, TH is Timing Hodoscope and SH represents the shower detector.} 
    \label{fig:bigbitearm}
\end{figure}

The BBS layout is shown in Fig.~\ref{fig:bigbitearm}. BBCal—the focus of this work—was designed, constructed, installed, commissioned, and calibrated for the BBS and demonstrated reliable performance throughout the SBS program. 

\section{Design}
\label{sec:calorimeter}

The BigBite calorimeter (BBCal) is a lead-glass electromagnetic calorimeter consisting of two parts: the preshower (PS) and shower (SH) layers. Together these layers provide a measurement of the total energy of the scattered electrons. The main purposes of BBCal for each experiment are the following:
\begin{itemize}
    \item Act as the main experiment trigger with good time resolution,
    \item Measure the total energy of electrons scattered from the target,
    \item Identify and differentiate incoming pions and electrons,
    \item Define a region-of-interest for fast and efficient track-finding in the GEM system.
\end{itemize}


\subsection{Preshower Detector} 
\label{PSdet}

The PS layer of BBCal was updated from its ``old" design used during JLab's $6\,$GeV era to include new lead-glass (LG) blocks (refurbished from the HERMES experiment \cite{AVAKIAN199869}) for better radiation hardness as well as to implement better magnetic field shielding in the form of mu-metal plates between each row of blocks and 0.25-inch thick soft iron plates in front of and behind the PS layer. The updated PS detector was constructed in the fall of 2020 at JLab and consists of 52 F101 LG blocks \cite{AVAKIAN1996155}, each with dimensions of $29.5\times9\times9\,\text{cm}^3$. Tab.~\ref{chemcomp_LG} provides details on the properties of these blocks. 

\begin{table}[h!]   
    \caption{\label{chemcomp_LG} Chemical composition and important properties of both types of LG blocks used in BBCal.}
    \centering
    \begin{tabular}{c | c c} 
        \hline \hline \vspace{-0.8em} \\
        LG Properties & F101 (PS) & TF1 (SH) \\
        \hline\vspace{-0.8em} \\
        Chemical Comp. & \multicolumn{2}{c}{Weight (\%)} \\
            \ce{Pb3O4} & 51.23 & 0.0 \\
            \ce{PbO} & 0.0 & 51.2 \\
            \ce{SiO2}  & 41.53 & 41.3 \\
            \ce{K2O}   & 7.0 & 7.0 \\
            \ce{As2O3} & 0.0 & 0.5 \\
            \ce{Ce}    & 0.2 & 0.0 \\ [0.5ex]
        \hline \vspace{-0.8em}\\
        Density           & \SI{3.86}{g/cm^{3}} & \SI{3.86}{g/cm^{3}}\\ 
        Refractive Index  & \SI{1.65}{} & \SI{1.65}{} \\
        Radiation Length  & \SI{2.78}{cm} & \SI{2.50}{cm}\\
        Moli\`ere Radius  & \SI{3.28}{cm} & \SI{3.50}{cm} \\ 
        Critical Energy   & \SI{17.97}{MeV} & \SI{15.00}{MeV} \\ [0.5ex]
        \hline\hline
    \end{tabular}
\end{table}

\begin{figure}[h!]
    \centering
    \includegraphics[width =0.95\linewidth]{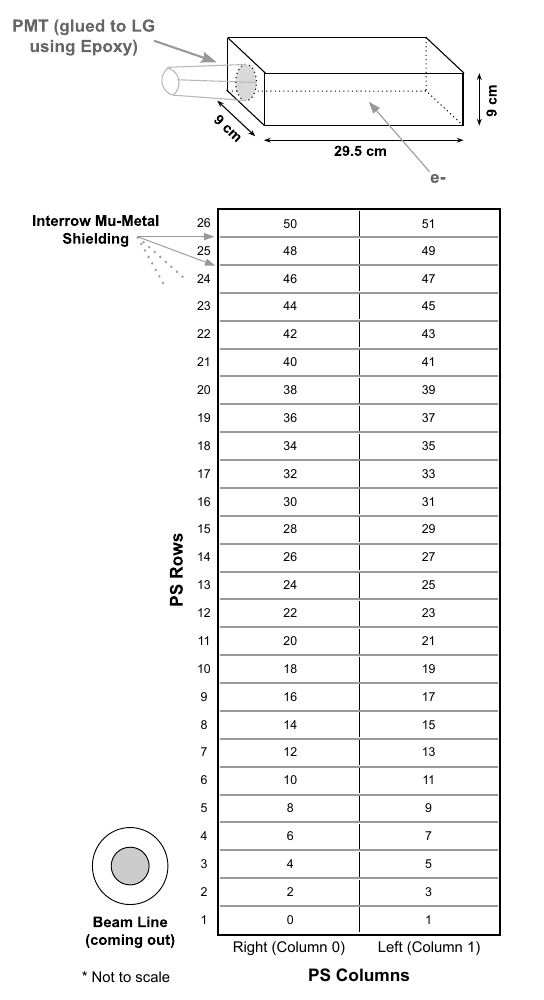}
    \caption{PS detector map (back view). A schematic of an individual PS module is included above the detector map. The Epoxy used to connect the PMTs to the LG blocks was UV-cured.}
    \label{fig:ps-sketch}
\end{figure}

Each LG block is attached to a Philips XP3461/PA photomultiplier tube (PMT) \cite{HERMES_cal} which reads out the Cherenkov radiation emitted by relativistic charged particles and secondary $e^+/e^-$ pairs produced in electromagnetic cascades. This combination of LG block and PMT serves as the basic unit of the PS layer. The LG blocks are arranged in 26 rows of two columns oriented perpendicular to the spectrometer axis, with the PMTs connected on the outer edge of each column. The sides of the the LG blocks on the inner edge of the columns were painted with EJ-510 reflective paint to reflect light back towards the PMT. As shown in Fig.~\ref{fig:ps-sketch}, the PMTs attached to column 0 (1) blocks are on the right (left) side of the PS, when looking downstream.

The PS detector frame is not light-tight because the sides remain open to allow passage for the high voltage and signal readout cables from the PMTs. Thus, before installation, each PS block needed to be individually wrapped in two layers:

\begin{enumerate}
    \item The inner layer is constructed from aluminized mylar, with the aluminized side facing outward. 
    \item The outer layer is constructed from black Tedlar film to provide optical isolation from any stray outside light. 
\end{enumerate}

\begin{figure}
    \centering
    \includegraphics[width= 0.95\linewidth]{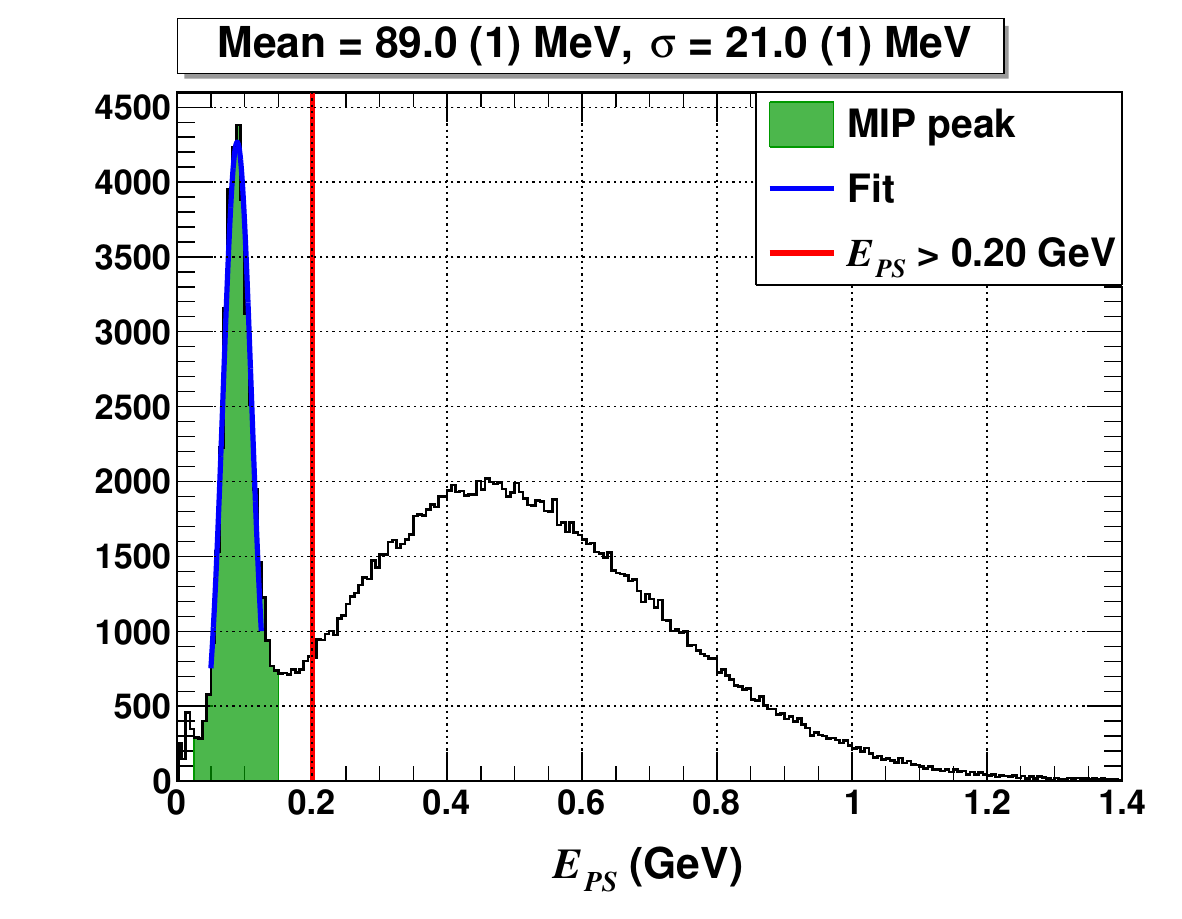}
    \caption{PS cluster energy distribution showing a prominent low-energy peak at approximately $89$ MeV, characteristic of MIPs. This feature enables effective rejection of pions from electrons using a simple threshold cut, indicated by the red vertical line. The data shown were obtained at $Q^2=4.5$~GeV$^2$ with $E_{\text{beam}}=4.0$~GeV during the E12-09-019 experiment and are typical of PS energy spectra obtained throughout the SBS program. For this setting, the scattered electron energies of interest ranged from 1.4-1.9~GeV. Figure adapted from \cite{Datta:2024vwq}.}
    \label{fig:pse}
\end{figure}

The orientation of the PS blocks requires scattered particles to traverse approximately 3 radiation lengths ($\sim9$~cm; see Tab.~\ref{chemcomp_LG}) of material, which is a sufficient thickness to initiate an electromagnetic shower and increase the electron signal well above the minimum ionizing particle (MIP) signal, while minimizing the probability for charged pions (and other hadrons) to generate large signals. Pions deposit a well-known, low energy of approximately 89~MeV in the PS, and this signature is readily identifiable in the PS cluster energy distribution, shown in Fig.~\ref{fig:pse}.

\subsection{Shower Detector} 
\label{SHdet}

\begin{figure}[h!]
    \centering
    \includegraphics[width=0.85\linewidth]{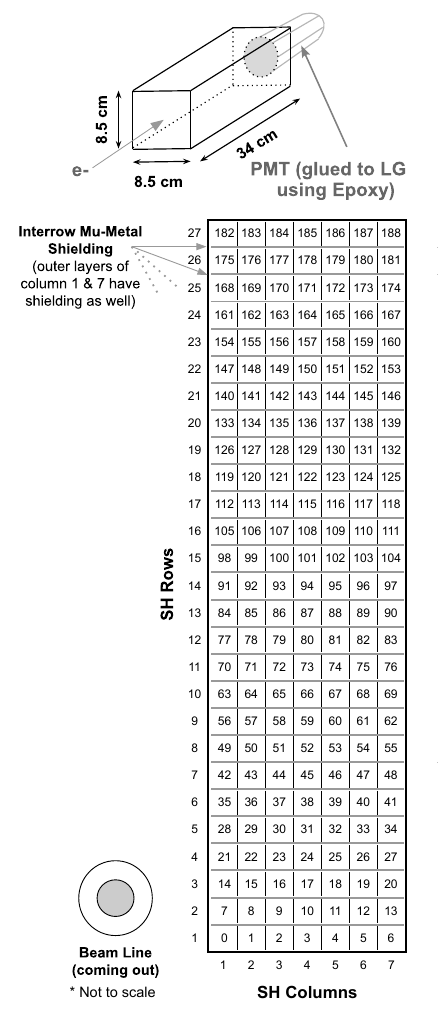}
    \caption{SH detector map (back view). A schematic of an individual SH module is included above the detector map. The Epoxy used to connect the PMTs to the LG blocks was UV-cured.}
    \label{fig:sh-sketch}
\end{figure}

The SH layer was also refurbished since its previous use at JLab. Similar to the PS, each module of the SH consists of one LG block and a corresponding PMT. The SH layer consists of 189 TF1 LG blocks \cite{article}\cite{LG_cheren} each with dimensions of $8.5\times8.5\times34\,\text{cm}^3$. Tab.~\ref{chemcomp_LG} provides details on the properties of the SH blocks. The blocks are stacked in 27 rows of 7 columns which face the incoming scattered particles and are oriented longitudinal to the spectrometer axis, as shown in Fig.~\ref{fig:sh-sketch}.

\begin{figure}[h]
    \centering
    \includegraphics[width= 0.9\linewidth]{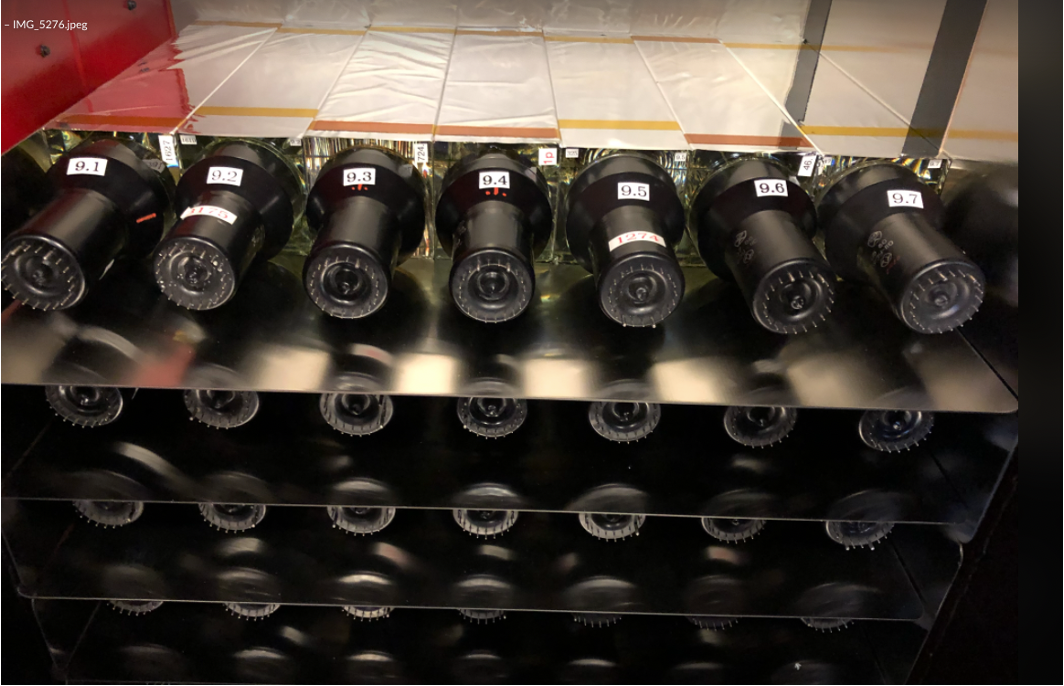}
    \caption{Sheets of mu-metal installed between each row of SH PMTs. This image is from the back of the SH layer during construction.}
    \label{fig:muMetal}
\end{figure}

The SH blocks use two types of PMTs: ITEP FEU-110 \cite{FEU-110} and Photonis XP5321B \cite{photonis}. The SH detector frame is light-tight, so the SH blocks are only wrapped in aluminized mylar. Similar to the PS, mu-metal sheets (alternating layers of 0.04~in and 0.05~in between SH layers and 0.05~in on the enclosure walls) were installed between each row as well as above and below the first and last row (see Fig. \ref{fig:muMetal}) in order to provide shielding against stray magnetic fields coming from the SBS and BBS dipole magnets.

The thickness of the SH blocks corresponds to 14 radiation lengths. Thus, the PS and SH together constitute approximately 17 radiation lengths, sufficient to fully contain the high-energy electrons of interest.

\section{Operation}
\label{sec:operation}

\subsection{Trigger and Data Acquisition}
\label{sec:DAQ}

\begin{figure*}[h]
    \centering
    \includegraphics[width= 0.9\linewidth]{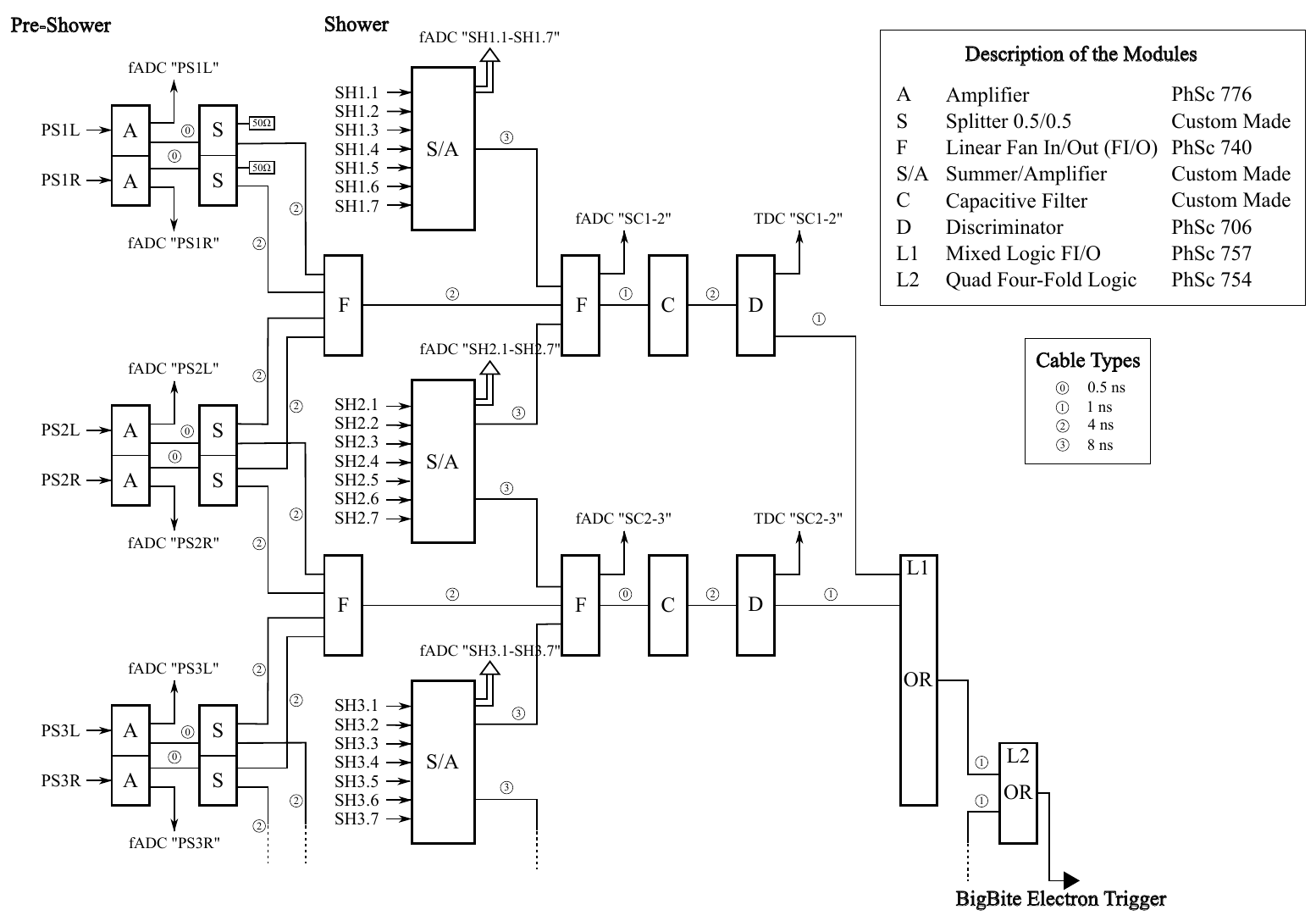}
    \caption{Schematic of a subset of the FE electronics adapted from \cite{Datta:2024vwq}.}
    \label{fig:FE}
\end{figure*}

BBCal was used to define the electron trigger for all the SBS experiments using BBS. The raw PMT signals from the PS and SH layers were amplified and split within the front-end (FE) electronics. One copy of the signals was sent to a flash Analog-to-Digital Converter (fADC 250) module for data acquisition \cite{FADC250}, and the other was used to form the electron trigger. The copies of PS and SH signals took slightly different paths through the FE electronics as shown in Fig.~\ref{fig:FE} and described below.

Raw PMT signals from the PS were first sent to a 2-output 10x Phillips Scientific (PhSc) amplifier model 776. The two amplified output copies were then processed separately:

\begin{enumerate}
    \item One copy was sent directly to a fADC 250 module for digitization.
    \item The other was sent to a custom-made 2-output passive splitter module from which two identical outputs were generated, each amplified by a factor of approximately five. These outputs were then routed into a linear fan-in/fan-out (LFI/O) module to sum signals from overlapping PS rows. These sums were finally sent to PhSc model 740 quad LFI/O modules for final trigger formation.
\end{enumerate}

Raw PMT signals from the SH were sent directly to custom-made Summer/Amplifier (S/A) modules and were split into two copies:
\begin{enumerate}
    \item One copy was amplified by a factor of approximately five before being sent directly to a fADC 250 module for digitization.
    \item The other copy was amplified by a factor of approximately 3.5 and was summed with amplified signals from the other six SH PMTs in the same row on the calorimeter. The summed output was then sent to the same PhSc model 740 quad LFI/O modules as the PS sums for final trigger formation.
\end{enumerate}

\subsubsection{Trigger Sums}
\label{trigger_sums}

In the LFI/O modules where both SH and PS signals were sent, the signals were combined into 25 trigger sums (labeled SC1-2, SC2-3, etc.) each comprising both SH and PS signals, as shown in Tab.~\ref{tab:bbcaltrigsum}. The inclusion of three SH rows instead of two in the 11 trigger sums formed by the middle rows of the SH and PS gives more weight to the events generated within the experimental acceptance. Additionally, such a design accounts for the slight mismatch in geometric alignment between SH rows and their corresponding PS rows.

\begin{table}[h!]
    \small 
    \centering
    \caption{ \label{tab:bbcaltrigsum} List of BBCal trigger sums formed by the SH and PS rows. Here, PS-1, PS-2, etc., represent the sums of the amplified signals coming from the left and right modules on the PS layer, and SH-1, SH-2, etc., represent the sums of the amplified signals coming from all the seven modules in one SH layer. SC 1-2, SC 2-3, etc., simply represent the different sums from corresponding PS and SH rows.}
    \begin{tabular}{ll}
    \hline\hline\vspace{-1.1em} \\ 
        Trigger Sums & Associated SH $\&$ PS Rows  \vspace{0.2em} \\ \hline \vspace{-1.1em} \\
        SC 1-2 & SH-1 + SH-2 + PS-1 + PS-2\\
        SC 2-3 & SH-2 + SH-3 + PS-2 + PS-3\\
        SC 3-4 & SH-3 + SH-4 + PS-3 + PS-4\\
        SC 4-5 & SH-4 + SH-5 + PS-4 + PS-5\\
        SC 5-6 & SH-5 + SH-6 + PS-5 + PS-6\\
        SC 6-7 & SH-6 + SH-7 + PS-6 + PS-7\\
        SC 7-8 & SH-7 + SH-8 + PS-7 + PS-8\\
        SC 8-9 & SH-8 + SH-9 + SH-10 + PS-8 + PS-9\\
        SC 9-10 & SH-9 + SH-10 + SH-11 + PS-9 + PS-10\\
        SC 10-11 & SH-10 + SH-11 + SH-12 + PS-10 + PS-11\\
        SC 11-12 & SH-11 + SH-12 + SH-13 + PS-11 + PS-12\\
        SC 12-13 & SH-12 + SH-13 + SH-14 + PS-12 + PS-13\\
        SC 13-14 & SH-13 + SH-14 + SH-15 + PS-13 + PS-14\\
        SC 14-15 & SH-14 + SH-15 + SH-16 + PS-14 + PS-15\\
        SC 15-16 & SH-15 + SH-16 + SH-17 + PS-15 + PS-16\\
        SC 16-17 & SH-16 + SH-17 + SH-18 + PS-16 + PS-17\\
        SC 17-18 & SH-17 + SH-18 + SH-19 + PS-17 + PS-18\\
        SC 18-19 & SH-18 + SH-19 + SH-20 + PS-18 + PS-19\\
        SC 19-20 & SH-20 + SH-21 + PS-19 + PS-20\\
        SC 20-21 & SH-21 + SH-22 + PS-20 + PS-21\\
        SC 21-22 & SH-22 + SH-23 + PS-21 + PS-22\\
        SC 22-23 & SH-23 + SH-24 + PS-22 + PS-23\\
        SC 23-24 & SH-24 + SH-25 + PS-23 + PS-24\\
        SC 24-25 & SH-25 + SH-26 + PS-24 + PS-25\\
        SC 25-26 & SH-26 + SH-27 + PS-25 + PS-26\\ \vspace{-1.2em} \\
    \hline\hline
    \end{tabular}
\end{table}

Two of the outputs from the quad LFI/O were used for the following purpose:
\begin{enumerate}
    \item One output was sent directly to a fADC 250 module as part of the trigger performance monitoring system.
    \item The other output was used to form the main electron trigger. 
\end{enumerate}

Each of these final LFI/O outputs were filtered using a high-pass filter to get rid of any DC offsets and baseline fluctuations. 
The filtered trigger sums were then processed by PhSc model 706 discriminators which were modified to allow remote threshold adjustment. The output pulse widths of the discriminators were kept constant at 40~ns throughout all the experiments.

\subsubsection{Threshold Determination} \label{sec:targetamp}

An optimal remotely adjustable threshold on the BBCal trigger was chosen such that the DAQ live time was maximized while the loss of physics events of interest was minimized. The minimum expected energy of the scattered electrons of interest for each experimental kinematic configuration was determined via realistic simulation, discussed in Sec.~\ref{sec:simulation}, and this value was used to define the threshold such that scattered electrons above this energy were recorded. However, in order to convert the energy (in MeV) to the threshold setting (in mV) at the discriminator level, we determined a threshold conversion factor, $Th_{CF}$, defined as follows:

\begin{equation} \label{ThCFeqn}
    Th_{CF}=C\times A_{Trig}
\end{equation}

\noindent where $A_{Trig}$ is the signal amplitude in units of mV at the trigger level\footnote{Trigger level here means the signal at the input of the quad LFI/O where the trigger sums are made.} after cosmic calibrations (see Sec.~\ref{subsec:coscal}) and $C$ is a constant in units of MeV$^{-1}$. Tab.~\ref{tab:CF} shows the conversion factors found for different kinematic settings during E12-09-019. An initial empirical value for $C$ was estimated using the known cosmic muon energy deposition per BBCal block, $72~$MeV (see Sec.~\ref{subsec:coscal}). Later, the value was fine-tuned by comparing with the rising edge of the BBCal cluster energy distribution obtained from beam-on-target data. 

\begin{table*}[]
    \centering 
    \caption{Scattered electron energy ($E'_e$), 4-momentum transfer ($Q^2$), beam energy ($E_{\text{beam}}$), and threshold conversion factor ($Th_{CF}$) for different E12-09-019 kinematic configurations.}
    \label{tab:CF}
    \begin{tabular}{cccccc}
        \hline\hline\vspace{-1em} \\ 
        Central $E'_{e}$ (GeV) & Minimum $E'_e$ (GeV) & Maximum $E'_e$ (GeV) & $Q^2$ (GeV/c)$^2$ & $E_{\text{beam}}$ (GeV) & $Th_{CF}$ (mV/MeV) \vspace{0.2em} \\ \hline \vspace{-1.1em} \\
        1.6 & 1.43 & 1.86 & 4.5 & 4.0 & 0.44 \\
        2.0 & 1.75 & 2.31 & 7.4 & 6.0 & 0.35 \\
        2.1 & 1.88 & 2.39 & 3.0 & 3.7 & 0.35 \\
        2.7 & 2.27 & 3.15 & 9.9 & 7.9 & 0.26 \\
        2.7 & 2.29 & 3.25 & 13.6 & 9.9 & 0.26 \\
        3.6 & 3.09 & 4.14 & 4.5 & 6.0 & 0.18 \\
        \hline\hline
    \end{tabular}
\end{table*}

We chose $A_{Trig}$ during calibrations based on the saturation level of the trigger electronics, specifically the saturation level of the S/A modules which is 200 mV. The saturation limit of the S/A modules was determined to be the limiting factor in defining the dynamic range of the entire analog electronics chain. We use this saturation value along with the maximum expected scattered electron energy from simulation, $E_e^{Max}$, to calculate an upper bound on our $A_{Trig}$ value:

\begin{equation} \label{trigEqn}
    A_{Trig}^{Max}\leq C_{elec}\times\frac{E_{dep}^{cos}}{E_e^{Max}}\times200\text{mV}
\end{equation}

\noindent where $C_{elec}$ is the factor due to the combined contributions of amplification and signal attenuation within our electronics, and $E_{dep}^{cos}$ is the average total energy deposited in a LG block by a cosmic muon.

Once Eqn.~\ref{trigEqn} was used to determine a maximum value for our $A_{Trig}$ setting, this value was then used along with Eqn.~\ref{ThCFeqn} to calculate the threshold conversion factor. As an example, in E12-09-019, the $Q^2=4.5~$(GeV/c)$^2$ kinematic configuration had an $A_{Trig}^{Max}$ of 10~mV and a $Th_{CF}$ of  0.18~mV/MeV.

\subsubsection{Signal Amplitude Mapping}
\label{sec:ampmap}

In order for the electron trigger to be stable and efficient, the PMT signal amplitudes needed to be matched at the trigger level, despite only having the ability to read the fADC signals using our DAQ system. Due to differences in amplification and attenuation between the two PMT signal copies which were split at the FE, there was a clear discrepancy between the signal amplitude at the trigger level and at the fADCs. Thus, it was vital to establish a map between these two signals.

\begin{figure}[h!]
    \centering
    \includegraphics[width=0.9\linewidth]{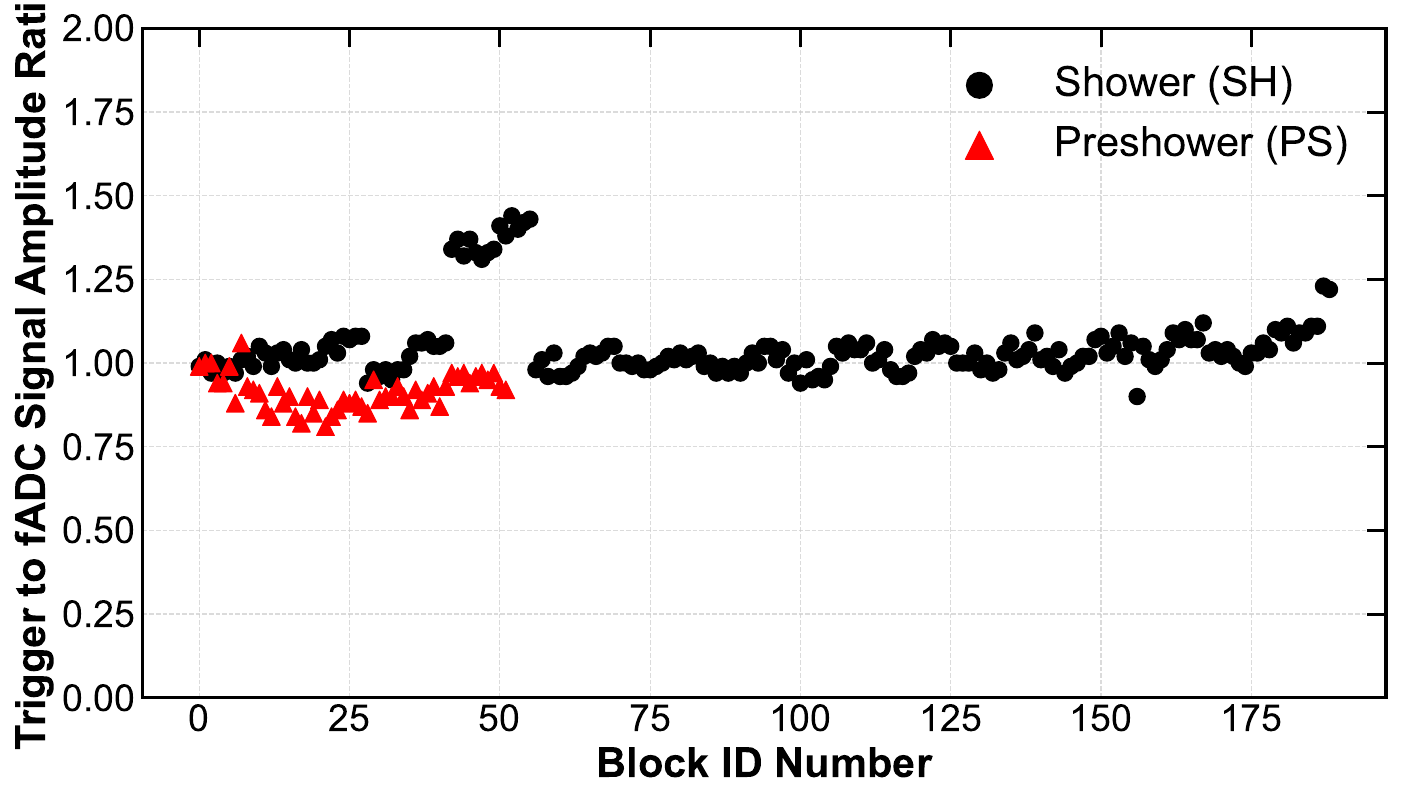}
    \caption{Ratios of signal amplitudes at the trigger over those at the fADC for all BBCal modules. Red triangular data points represent PS ratios, and black circular points represent SH ratios.}
    \label{fig:trig2fadc}
\end{figure}

A mapping needed to be defined for each channel in order to account for the gain variation across the S/A modules. A procedure was developed to give pulses with known amplitudes as an input to each S/A module and record the resulting signal amplitude as an output at the trigger level with an oscilloscope. Then, the corresponding signal was recorded at the fADC. These two signals were compared, and ratios were made for each channel which constitute the mapping between trigger level and fADC. As can be seen in Fig.~\ref{fig:trig2fadc} the ratios were approximately equal to 1 for almost all of the modules, but some channels saw a significant difference in the signals which was properly accounted for by applying these ratios. Thus, the mapping was used to match signal amplitudes at the trigger level.

\subsection{Cosmic Ray Calibrations}
\label{subsec:coscal}

Initial energy calibrations for the calorimeter were performed using cosmic ray data at the beginning of every experimental configuration. Later, a more sophisticated method was used to fine-tune the energy calibration coefficients using beam-on-target data described in Sec.~\ref{subsec:beamcal}. 


Due to their nature as MIPs, cosmic-ray muons deposit a relatively small and well-defined amount of energy in the calorimeter’s LG blocks. For a single vertical muon traversing one block (in either the SH or PS), the effective average energy deposition is approximately 72~MeV\footnote{After an initial value was calculated using an ideal circuit, it was then fine-tuned using BBCal calibrated beam-on-target data.}. Our initial energy calibration exploited this feature to equalize the ADC response across blocks by gain-matching the PMT voltages. To ensure a uniform calibration sample, only vertical cosmic rays were selected. The event selection imposed a ``verticality cut,” defined as follows:

\begin{itemize}
    \item For the SH, an event was accepted if the four vertical neighbors of a given block (the two above and two below) showed good signals, while its two horizontal neighbors did not exceed threshold.
    \item For the PS, an analogous cut was applied, requiring good signals only in the four vertical neighbors. 
\end{itemize}

After applying this selection, signal amplitude distributions were fitted to extract peak positions. These peak values were then aligned to a common reference by adjusting the PMT high voltages.

\begin{eqnarray}
    \label{eqn:hv}
    HV_{new} = HV_{old} \left( \frac{A^{set}_{Trig}}{V_{old}} \right)^{\frac{1}{\alpha}}
\end{eqnarray}
\noindent where:

\vspace{-0.5em}
\begin{align*}
    \alpha &\equiv \text{PMT gain exponent} \\
    HV_{old} &\equiv \text{PMT HV before calibration} \\
    HV_{new} &\equiv \text{desired calibrated PMT HV value} \\
    V_{old} &\equiv \text{signal amplitude before calibration} \\
    A^{set}_{Trig} &\equiv \text{desired signal amplitude after calibration (Sec.~\ref{sec:targetamp})} \nonumber
\end{align*}

\begin{figure}[h!]
     \centering
     \begin{subfigure}[b]{0.45\textwidth}
         \centering
         \includegraphics[width= \textwidth]{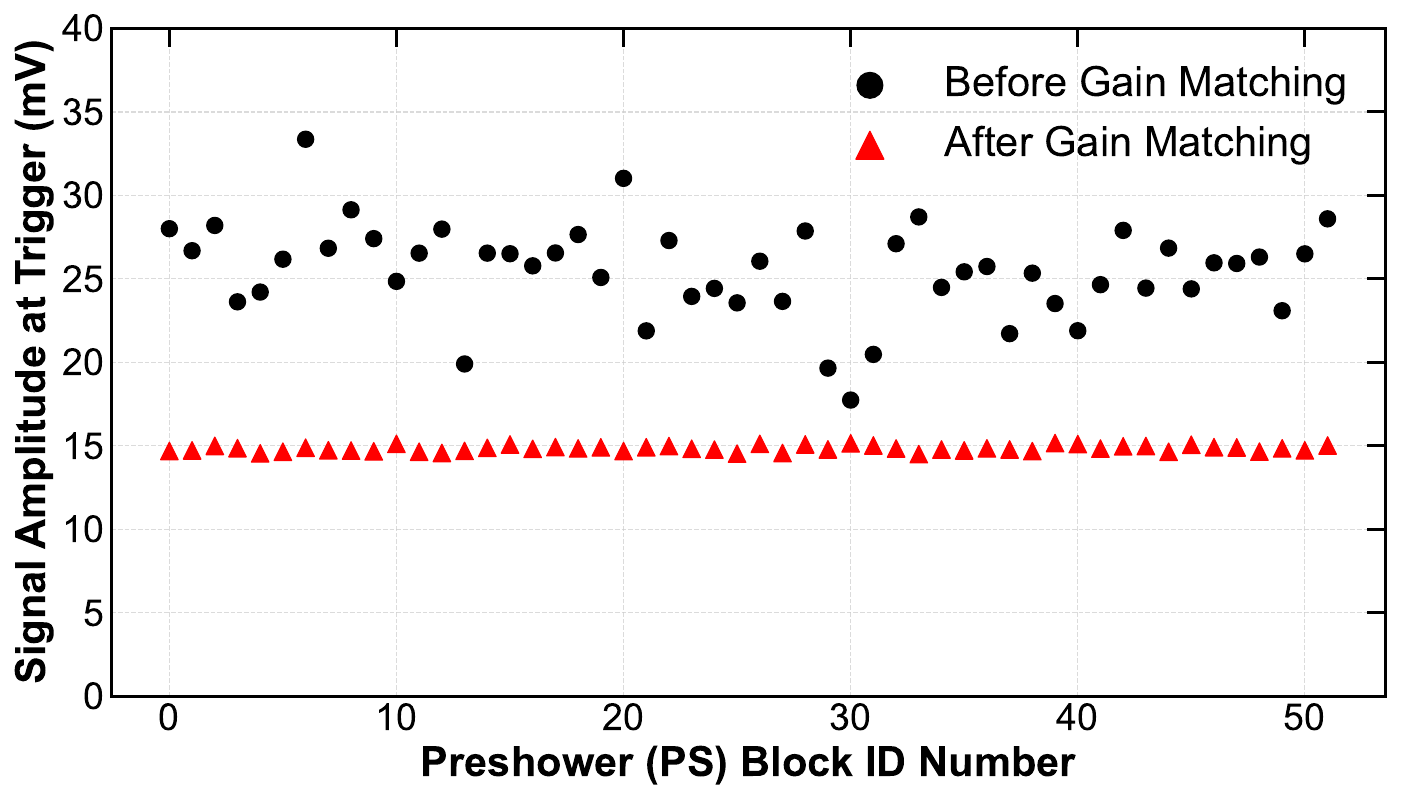}
         \caption{}
         \label{sfig:PS_gain_match}
     \end{subfigure}
     \bigskip
     \begin{subfigure}[b]{0.45\textwidth}
         \centering
         \includegraphics[width= \textwidth]{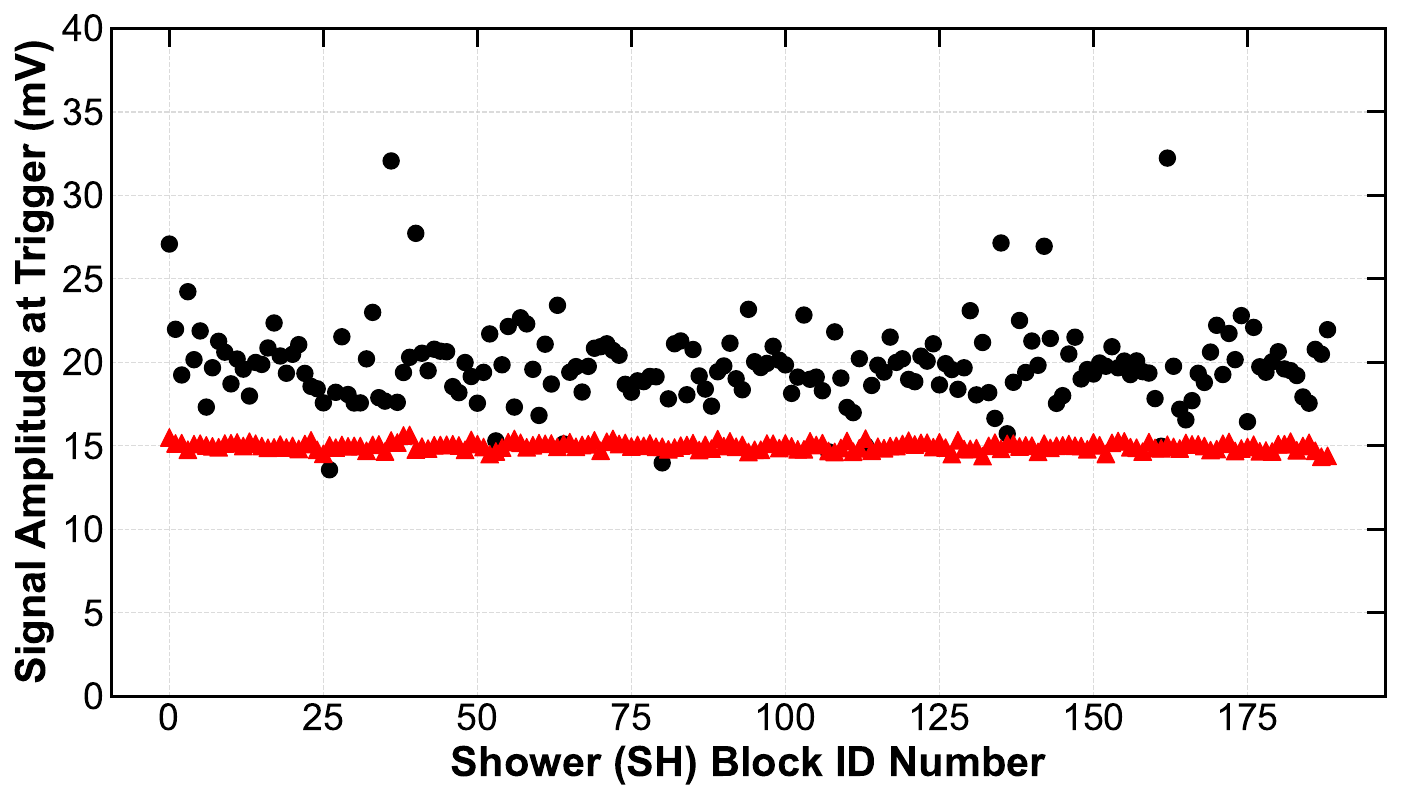}
         \caption{}
         \label{sfig:SH_gain_match}
     \end{subfigure}
     \caption{Effect of gain-matching on signal amplitudes for PS (a) and SH (b) blocks. Black circular data points represent amplitudes before gain-matching, and red triangular data points represent amplitudes after gain-matching.}
     \label{fig:bbcalgainmatch}
\end{figure}

\noindent The effect of this gain-matching using cosmic data can be seen in Fig.~\ref{fig:bbcalgainmatch}.

\begin{figure}[h!]
    \centering
    \includegraphics[width= 0.95\linewidth]{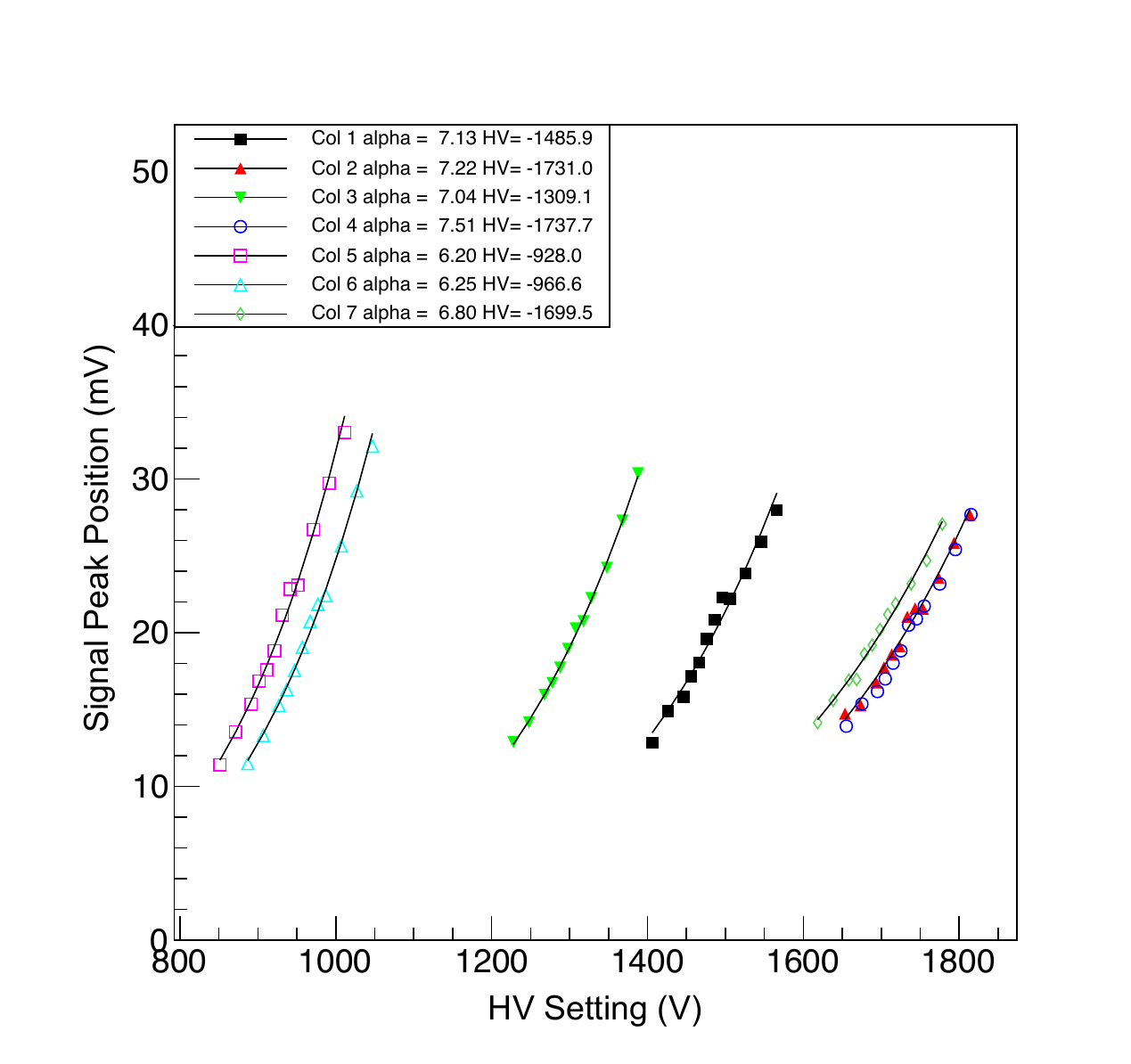}
    \caption{Example results of an HV scan done in January 2024. The channels plotted are the seven PMTs in SH row 4. The HV settings were varied in 10-V increments within a $\pm 50$-V range centered at each channel's nominal HV setting. The legend lists the nominal HV settings and the $\alpha$ values resulting from the fits using Eqn.~\ref{eqn:hv} for each PMT.}
    \label{fig:HV_scan}
\end{figure}

We carried out several High Voltage (HV) scans to determine the gain exponent $\alpha$ for each PMT. This involved taking cosmic data using several HV settings covering the entire operational range of the PMTs. Then, the plot of peak position vs HV setting for each PMT was fit using a function of the same form as Eqn.~\ref{eqn:hv}, and this fit was used to extract the corresponding $\alpha$, as shown in Fig. \ref{fig:HV_scan}. Importantly, an HV scan was done prior to any beam-on-target data taking, and PMTs that were performing poorly were replaced. The $\alpha$ exponents remained the same for all of the PMTs throughout E12-09-019 and E12-09-016, but they were updated before E12-17-004 to account for some deteriorating PMT performance.

\subsubsection{Magnetic Field Mitigation}

\begin{figure*}[h!]
    \begin{subfigure}{0.5\textwidth}
        \centering
        \includegraphics[width= \textwidth]{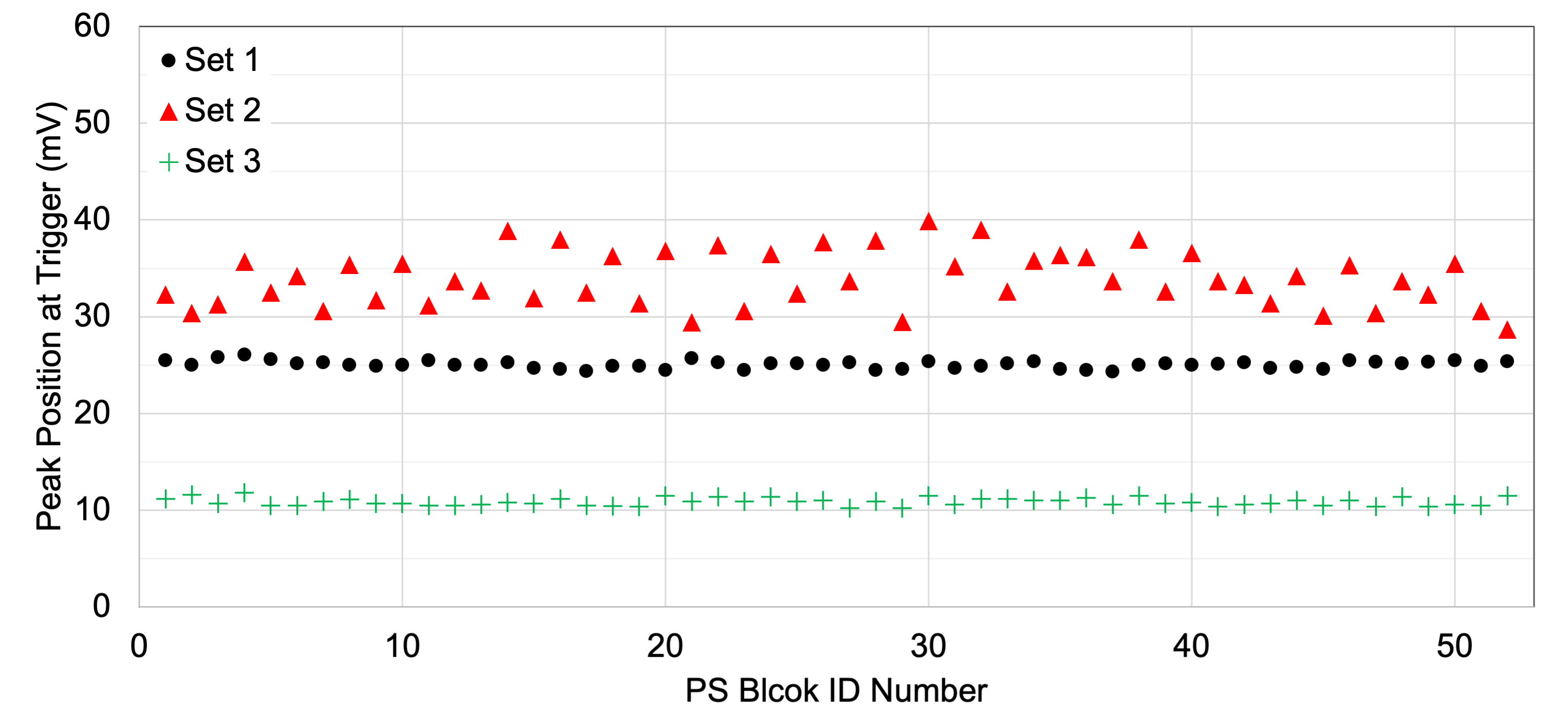}
        \caption{}
        \label{fig:PSfield}
    \end{subfigure}
    \begin{subfigure}{0.5\textwidth}
        \centering
        \includegraphics[width= \textwidth]{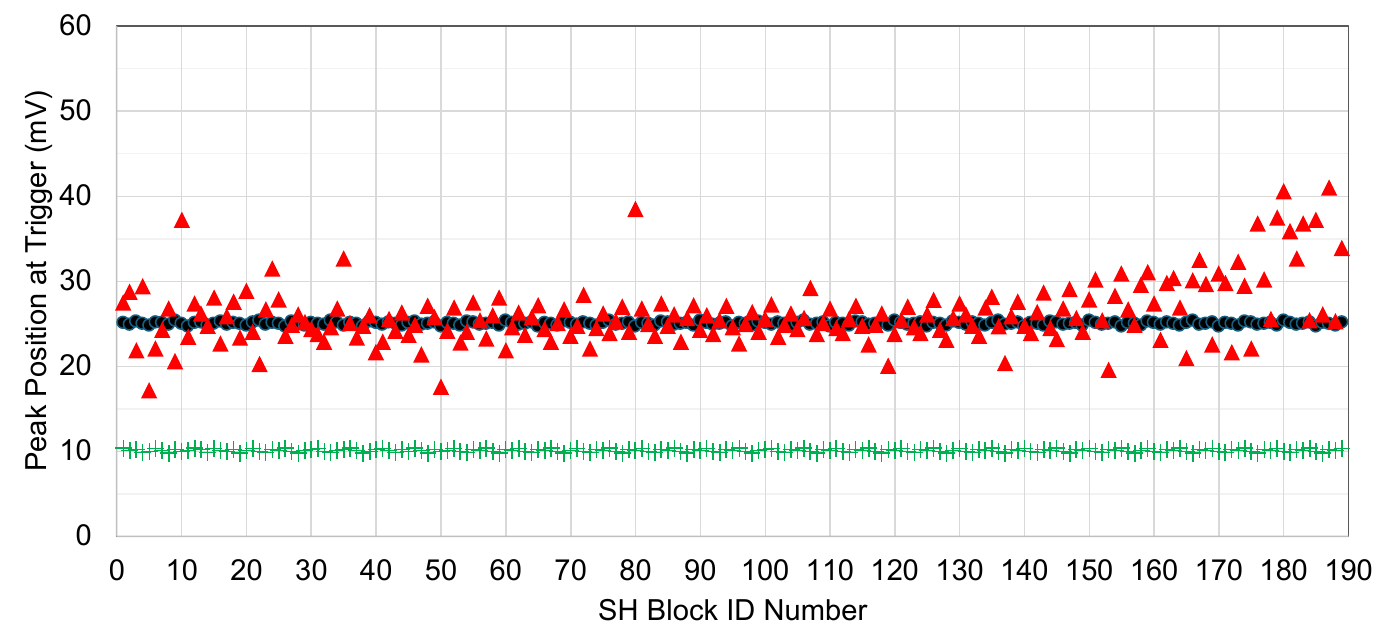}
        \caption{}
        \label{fig:SHfield}
    \end{subfigure}
    \caption{Mitigation of PS (a) and SH (b) PMT gain shifts induced by stray magnetic fields from the SBS and BBS dipole magnets ($\approx 50$~G near the outside of the PS and SH detector housing on the side closer to the beam line). Each data set shows the peak position of the cosmic-ray signal amplitude distribution as a function of block ID. Set 1: data from a cosmic-ray run taken with both magnets off, following initial gain-matching of the PMTs. Set 2: data from a run taken with both magnets on, demonstrating the degradation of the gain-matching due to fringe fields. Set 3: data after re-gain-matching using the magnet-on cosmic-ray run, yielding uniform PS and SH amplitudes of approximately \SI{10}{mV}. The target amplitude was chosen based on kinematic considerations to avoid saturation of the S/A modules.}
    \label{fig:mit}
\end{figure*}

Calibration of the electron trigger critically depends on precise gain-matching of the BBCal photomultiplier tubes (PMTs). Shifts in the PMT gains adversely affect the calibration, resulting in a biased and less efficient electron trigger. During the commissioning of experiment E12-09-019, significant shifts in the BBCal PMT gains were observed when changing the magnetic field settings. These shifts were attributable to the unexpectedly large fringe magnetic fields from the nearby SBS magnet, with a smaller effect from the BBS magnet, that were not completely shielded by the detector and PMT housing ($\approx 50$~G near the outside of the PS and SH detector housing on the side closer to the beam line.). Fig.~\ref{fig:mit} illustrates the impact of these gain variations on the PS and SH signal amplitudes, as observed in cosmic ray data.


In order to address this issue and continue with experimental data taking, the following plan was executed:

\begin{enumerate}
    \item Once the BBS and SBS arms were in their final positions for a given configuration, cosmic data were taken with no magnetic field.
    \item These data were used to gain-match the PMTs to a very high target signal amplitude ($\geq~$25 mV) so that no PMT signals were lost when the magnetic field was turned on.
    \item The BBS and SBS magnets were turned on at the strength needed for data taking, and more cosmic data was taken. The PMT gain-matching done during this process yielded HV settings that account for the fringe field.
\end{enumerate}

This procedure was repeated whenever there was a change in the experimental configuration. To avoid loss of beam time, a semi-automated analysis framework \cite{bbcalreplay} was developed that executes the entire workflow in less than 30 minutes. The system was designed to be sufficiently robust and user-friendly to allow opportunistic execution by non-expert shift personnel, thereby enabling rapid and reliable recalibration during routine operations.

\subsection{High Voltage Monitoring}
\label{sec:HVmon}

The HV distribution system for the BBCal PMTs consisted of two LeCroy 1458 HV crates with built-in Raspberry Pi (RPI)-based software controls and 21 LeCroy 1461N HV cards. The HV crates were installed in a rack, one on top of the other, in a shielded bunker in the experimental hall to avoid radiation damage during data taking. Each LeCroy 1458 HV crate can hold 16 type 1461N HV cards, each equipped with twelve HV supply channels. Such a design necessitates a total of twenty-one HV cards to be installed in two HV crates to accommodate all 241 BBCal PMTs.

The RPIs allow remote modifications to the HV settings of each individual channel by connecting to Jefferson Lab's Experimental Physics Industrial Control System (EPICS)\cite{Dalesio:1994qp}, and saving the readouts to an archive. This system also includes online monitoring of voltage and current read-backs for each PMT channel, and trip limits were set that will alarm when exceeded. The ability to continuously monitor and quickly and remotely modify the HV settings was vital for all the experiments.

\subsection{Online Monitoring}

For real-time data quality assurance, a fraction of the incoming data were decoded, reconstructed, and plotted in real time using the custom event reconstruction library~\cite{SBSoffline} mentioned in Sec.~\ref{sec:simulation}. A standard set of histograms of raw and reconstructed data distributions for each detector subsystem were generated and inspected by experiment shift crews to monitor detector performance, stability and data quality, identifying and debugging any issues that might arise in real time. Low-level raw signal information from the PS and SH included count rates, pedestal values, and ADC time and amplitude spectra for each block. High-level reconstructed quantities included cluster size, multiplicity, energy, and position distributions, correlations between SH and PS cluster energies, the ratio $E/p$ of the total shower energy measured by BBCal and the reconstructed track momentum, and correlations with detectors in the SBS arm, such as HCal. 

Various slow control parameters were also continuously monitored and archived by EPICS and periodically inserted into the raw data stream. For BBCal these included the PMT HV setpoint, readback, and current information (see Sec.~\ref{sec:HVmon}) and trigger threshold setpoint and readback values. Copies of the individual trigger sums and the final trigger signal were sent to scaler modules for real-time rate monitoring, and to various Time-to-Digital Converter (TDC) boards for performance monitoring, and to provide a common timing reference for all detector subsystems using TDC readout.

As mentioned in Sec.~\ref{trigger_sums}, the analog trigger sums (shown in Tab.~\ref{tab:bbcaltrigsum}) were also continuously monitored and read out via fADCs for additional cross checks on data quality, gain stability, and trigger efficiency. Trigger diagnostics included pedestal, ADC time, signal amplitude, and TDC time for each trigger sum, as well as correlations between the analog trigger sums and the sums of the fADC integrals over all the individual PMTs contained in each sum. The latter comparison served as a consistency check of the amplitude ratios between analog signals at the input of the trigger summing modules and at the fADC readout boards mentioned in Sec.~\ref{sec:ampmap}. 




\section{Data Analysis}
\label{sec:datana}

In this section, we describe the clustering algorithm used by BBCal and the calibrations of BBCal energy and timing reconstruction needed to achieve the design performance.

\subsection{Cluster Formation} \label{sec:clus}

The raw data from BBCal consists of 25 4-ns fADC waveform samples for all channels in every triggered event. After unpacking the raw data, the pedestal is estimated by averaging four voltage samples from the beginning and end of the waveform, and the estimated pedestals are then subtracted from all waveform samples. After pedestal subtraction, the threshold crossing sample is located, and for every channel with a signal above the user-configurable threshold (typically 5~mV) for pulse-finding in the waveform, the pulse amplitude, charge integral, and leading edge time are reconstructed. The charge integral is then converted to an energy deposit using a gain factor resulting from the calibration described in Sec.~\ref{subsec:beamcal}, generating a list of ``good” hits with above-threshold energy deposits. This list of BBCal blocks with good hits is then passed to the clustering algorithm described below.

\begin{figure}[h!]
    \centering
    \includegraphics[width= 0.95\linewidth]{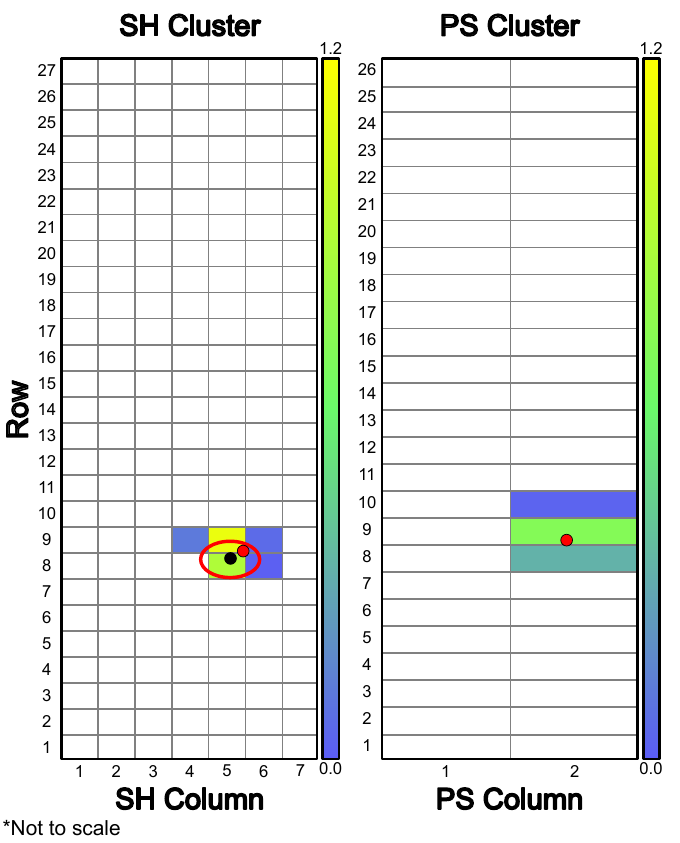}
    \caption{Visualization of the highest-energy BBCal cluster for an example event from E12-09-019. In both plots, the color scale indicates the energy deposited in each block in GeV. Colored blocks correspond to those passing the position, time, and energy criteria of the clustering algorithm. In the left plot, the red ellipse denotes the tracking system’s search region. The black filled circle indicates the SH cluster centroid, and the red filled circles represent the reconstructed electron track positions projected onto SH (left) and PS (right).}
    \label{fig:clusVis}
\end{figure}

SH and PS hits sufficiently close together in space and time yield individual SH and PS clusters. The combination of these clusters for a given event is defined as the BBCal cluster. Fig.~\ref{fig:clusVis} shows example clustering results from a single event with a high-energy electron track producing a shower in BBCal. The total energy deposition in these BBCal clusters is a measure of the scattered electron energy. 

Clusters encompass the entire EM shower, and multiple criteria help ensure that, to the greatest possible extent, each cluster contains only one detected particle and that we select the ``best" cluster for further analysis. For each triggered event, all hits that pass a minimum energy threshold are added to arrays for the SH and PS storing the position, time, energy, and block index of each hit, and these hits are then sorted by energy in descending order.


\begin{figure*}[ht!]
    \centering
    \begin{subfigure}{0.48\textwidth}
        \centering
        \includegraphics[width= \textwidth]{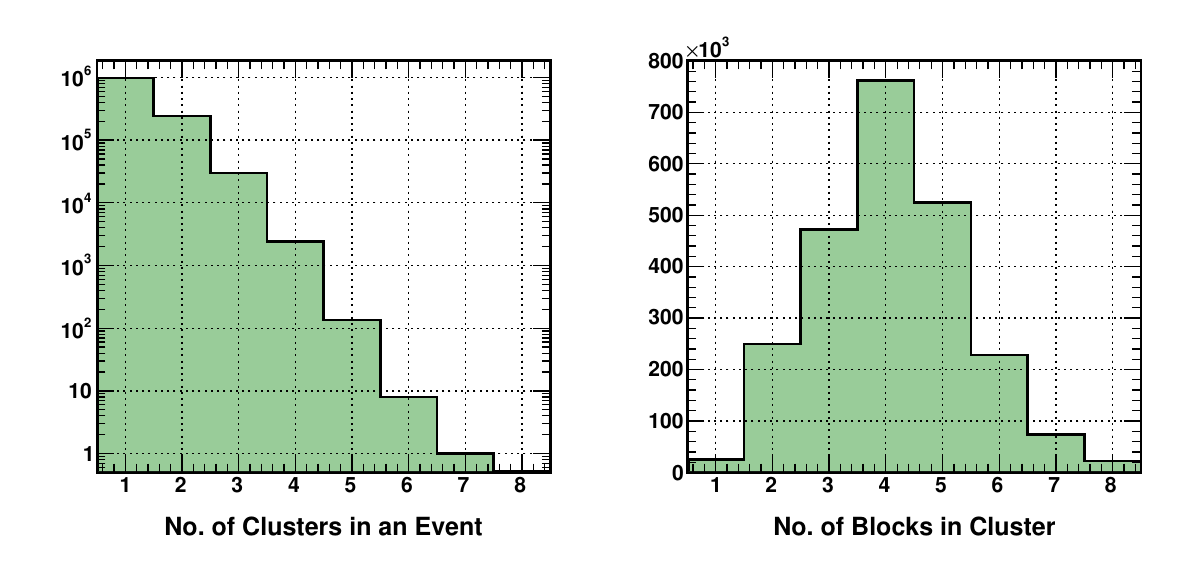}
        \caption{}
        \label{fig:SHclus}
    \end{subfigure}
    \hfill 
    \begin{subfigure}{0.48\textwidth}
        \centering
        \includegraphics[width= \textwidth]{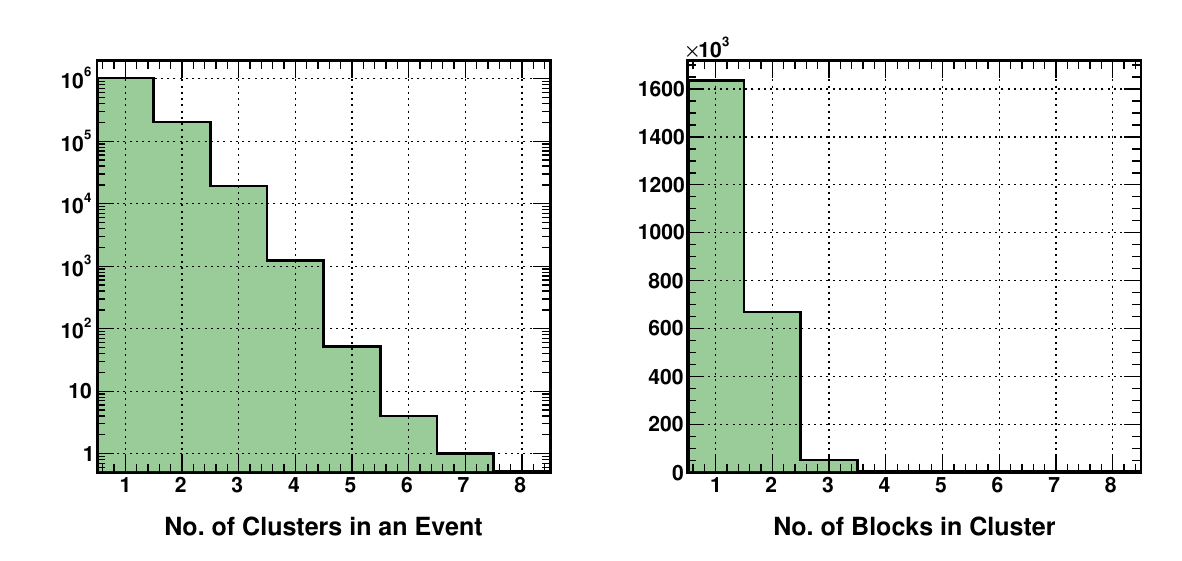}
        \caption{}
        \label{fig:PSclus}
    \end{subfigure}
    \caption{Cluster multiplicity and size distributions for SH (a) and PS (b) from E12-09-019, for the kinematics with $E = 6.0$~GeV and $Q^2 = 4.5$ GeV$^2$ at a beam current of 8~$\mu$A on the 15-cm liquid hydrogen target, corresponding to a luminosity of $3.2 \times 10^{37}\ e^-/\text{s} \cdot p/\text{cm}^2$. For this setting, BigBite was at a central angle of 26.5 degrees at a distance of 2 m from the target center to the front face of the iron yoke. The (quasi-)elastically scattered electron energy ranged from 3.1-4.1~GeV (see Tab.~\ref{tab:CF}), while the energy threshold for SH (PS) cluster formation was 300 (10) MeV. All distributions are shown for identified electron events with PS energy deposit above 0.2 GeV and reconstructed track momentum $p > 2.9$ GeV.}
    \label{fig:clusMult}
\end{figure*}


\subsubsection{SH Clusters}

The first block in the SH hit array has the greatest energy deposition and is used as the seed for the first cluster if its energy exceeds a user-configurable threshold, typically chosen to give high efficiency for the electrons of interest while excluding unwanted low-energy particles. The cluster is then formed using the ``island" algorithm described below. 


First, a new cluster object containing the seed block is created and added to the array of clusters for the event. Then, the seed block and any subsequent blocks added to the cluster are deleted from the unused hit array (``unused" meaning ``not yet added to any existing cluster") to avoid double counting. After removal of the seed block, remaining blocks in the unused hit array are compared one-by-one to all blocks that have already been added to the cluster in terms of the following quantities:

\begin{equation}
    r_{ji}^2\equiv(x_i-x_j)^2+(y_i-y_j)^2 \;\; , \;\; t_j\equiv |t_j^{ADC}-t_0^{ADC}| \nonumber
\end{equation}
where $(x_i,y_i)$ and $(x_j,y_j)$ are, respectively, the center coordinates of the $i$th block already added to the cluster and the $j$th unused block. $t_j^{ADC}$ and $t_0^{ADC}$ are, respectively, the ADC times (see Sec.~\ref{TimingSec}) of the $j$th block and the seed block. In order for a block to be added to the cluster array, the following criteria must be met:
\begin{equation} \label{clusterLims}
    r_{ji}^2\leq r_{max}^2 \;\;\; , \;\;\; t_j\leq t_{max}
\end{equation}
In other words, as blocks are added to the cluster and removed from the unused blocks array, remaining unused blocks are compared to every block that has been previously added to the cluster, and if an unused block is within a radius $r_{max}$ of any block already in the cluster and in time with the seed block (which generally has the best timing resolution), it is added to the cluster. This process continues until no more neighboring blocks are found satisfying the squared-distance and time criteria of Eqn.~\eqref{clusterLims}. In this way, clusters are allowed to grow to arbitrary size in any direction, essentially grouping all contiguous blocks whose hits are sufficiently close in time to the seed into clusters. At the end of this process, the centroid position of each SH cluster is reconstructed using a simple energy-weighted average of the center positions of all the blocks in the cluster.

For all experiments using BBCal, $r_{max}$ was chosen to be 15~cm. Given the SH block size and layout, this value accommodates nearest horizontal, vertical, and diagonal neighbors. $t_{max}$ was conservatively chosen to be 10~ns, and the rationale for this choice is described in more detail in Sec.~\ref{TimingSec}. Finally, the total energy of the SH cluster must exceed a user-configurable minimum energy threshold in order to be added to the array of clusters for each event. This cluster sum threshold is applied separately from the cluster seed threshold applied to the highest-energy block and is generally chosen to give high efficiency for the events of interest while suppressing unwanted low-energy events. 


After each iteration of cluster-finding, the highest-energy remaining block (if any) is used to seed a new cluster, and the process above is repeated until no new clusters are found passing both cluster seed and cluster sum energy thresholds.


\subsubsection{PS Clusters}

 After the SH clustering is finished, the PS data are searched for hits matching the SH clusters in position and time. Similar to the method of creating SH clusters, we start with an unused hit array of PS blocks for a given triggered event. 
 The energy-weighted centroid position of each SH cluster is projected onto the PS layer and compared to the center of each PS block. If both the vertical and horizontal distances between this cluster centroid position and a given PS block's center are less than 15~cm and 20~cm respectively, then the PS block is a position match to the SH cluster. 
 

The ADC time of each PS block is also compared to the ADC time of the SH cluster, and if these times are within 10~ns of each other, then the PS block is a timing match to the SH cluster. The ADC time of the resulting PS cluster is 
that of the position-matched PS block with the highest energy. Once a block is added to a cluster, it is removed from the unused hit array. 



\subsubsection{BBCal (SH+PS) Clusters}
We define the ``best" cluster in each event as that with the highest combined PS and SH energy. 
These best clusters are used to define a region of interest for the tracking algorithm, as mentioned in Sec.~\ref{sec:calorimeter} and illustrated in Fig.~\ref{fig:clusVis}. Representative examples of SH and PS cluster size and multiplicity distributions are shown in Fig.~\ref{fig:clusMult}.  
The cluster size distributions are shown for the best cluster. The multiplicity distributions include all SH clusters found by the aforementioned algorithm and all PS clusters matched with any SH cluster in position and time\footnote{The requirement of a matching shower cluster means that the PS multiplicity shown in Fig.~\ref{fig:PSclus} is always, by definition, less than or equal to the SH cluster multiplicity.} 

The cluster sum threshold used by the analysis shown in  Fig.~\ref{fig:clusMult} was 300 (10)~MeV for the SH (PS). 
This is significantly lower than the hardware trigger threshold, which was approximately 2.6~GeV for the setting in question. Roughly 24\% of all triggered events at this setting have two or more SH clusters. Under ordinary running conditions, the vast majority of secondary SH clusters represent accidental coincidences unrelated to the primary trigger. The fraction of true multi-particle coincidences in the SH resulting from the same collision during the same beam bunch crossing is on the order of 1\% of all multi-cluster events. As such, the SH cluster multiplicity distribution is, to a very good approximation, independent of the event topology for a given threshold. 
Given the effective fADC time window of 68~ns\footnote{25 fADC samples (100 ns) minus the four at each end of the waveform used for pedestal calculation gives 68 ns.}, the 24\% multi-cluster probability implies a single-particle rate in the entire BBCal of roughly 4 MHz\footnote{The \textit{triggered} event rate above the 2.6-GeV hardware threshold was roughly 2 kHz for the setting shown in Fig.~\ref{fig:clusMult}} above the 300-MeV software threshold.



\subsection{Calibrations}



\subsubsection{Energy}
\label{subsec:beamcal}


As described in Sec.~\ref{subsec:coscal}, rough gain-matching and energy calibrations using cosmic rays were performed prior to data taking for every experimental configuration. For offline event reconstruction and physics analysis, the BBCal gain coefficients were calibrated using electrons of known momentum scattered from hydrogen (and other targets when necessary). The momentum of the scattered electrons was reconstructed from tracking results and the BBS magnetic field map, with a typical resolution $\sigma_p/p\approx1-1.5\%$. Track momentum reconstruction was calibrated at each setting using elastic $H(e,e'p)$ events, for which the kinematics are over-determined by energy and momentum conservation.

\begin{figure}[h!]
    \centering
    \includegraphics[width= 0.9\linewidth]{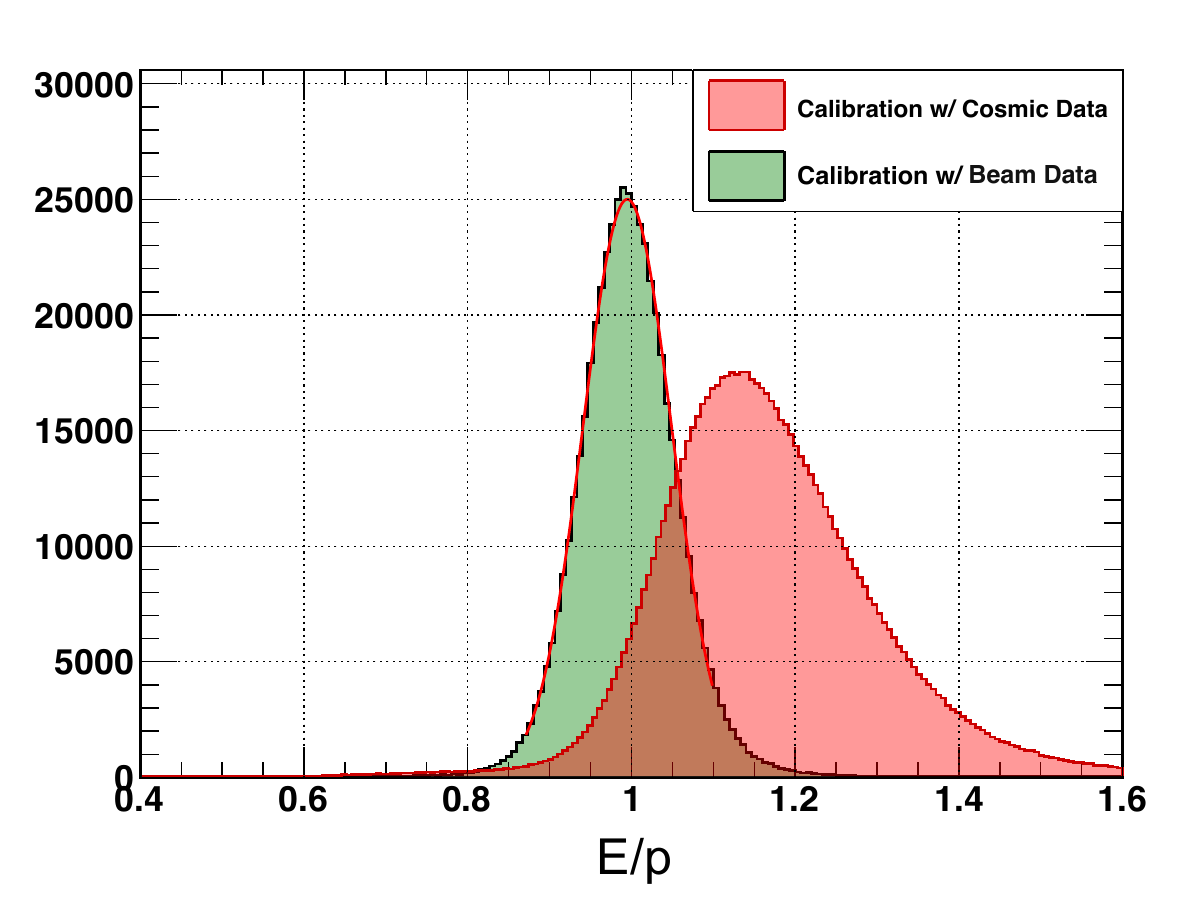}
    \caption{An example an $E/p$ ratio distribution from E12-09-019. The red (green) curve represents data reconstructed using the cosmic ray (beam)  calibration. Figure adapted from \cite{Datta:2024vwq}.}
    \label{fig:calib-compare}
\end{figure}

The quality of the energy calibration was confirmed using the ratio of $E$, the total calibrated energy deposited in BBCal, and $p$, the reconstructed electron track momentum. This $E/p$ ratio should ideally be one for the ultra-relativistic electrons of interest. Fig.~\ref{fig:calib-compare} shows that the energy reconstructed using the cosmic ray calibration was not uniform, as its $E/p$ distribution is asymmetric, peaking around 1.1, and significantly wider than the calibrated $E/p$ spectrum. The cosmic ray calibration was sufficient for rapid, convenient gain-matching and trigger optimization but suboptimal for physics analysis. The calibrated $E/p$ spectrum is Gaussian and peaks at 1 with a typical standard deviation of approximately 6\%. 

\begin{figure*}[h!]
    \centering
    \begin{subfigure}{0.8\linewidth}
        \centering
        \includegraphics[width= 0.9\linewidth]{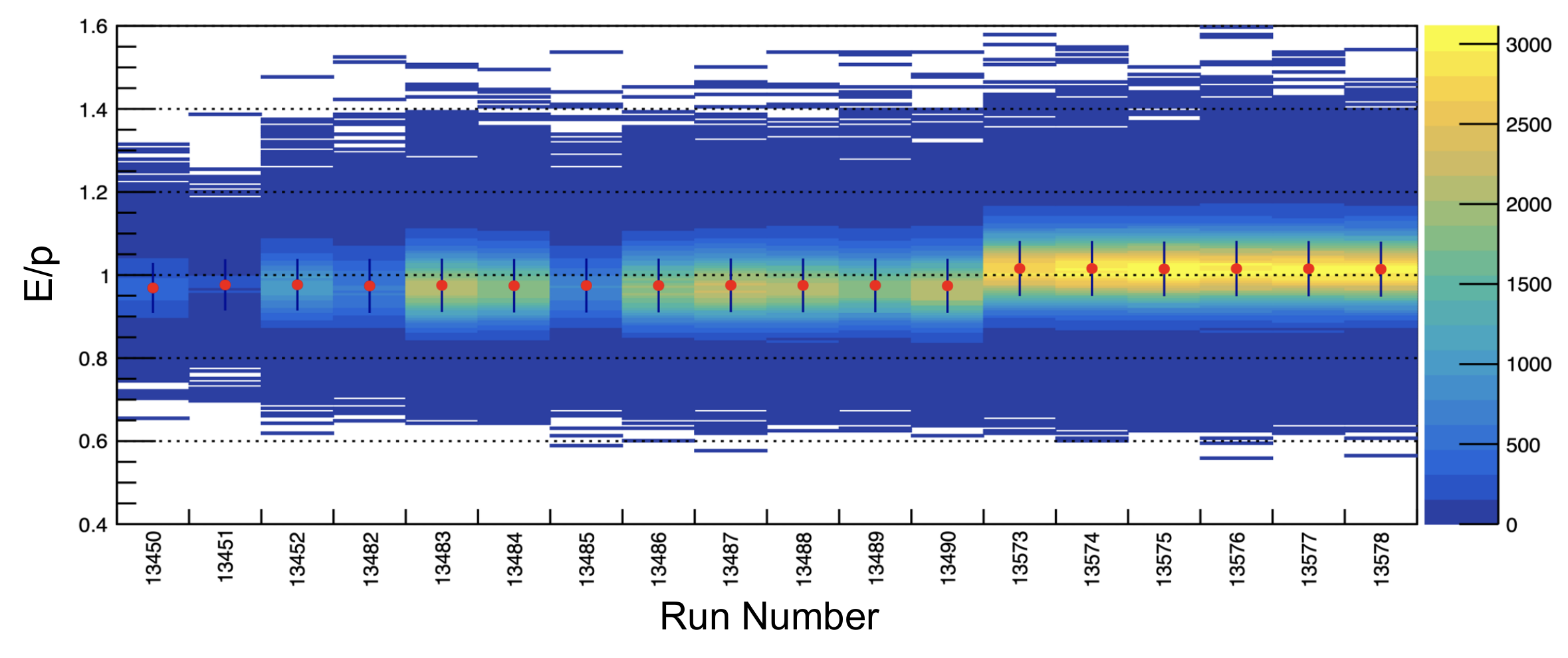}
        \caption{}
        \label{fig:sub-deviation}
    \end{subfigure}    
    \begin{subfigure}{0.47\textwidth}
        \centering
        \includegraphics[width= \textwidth]{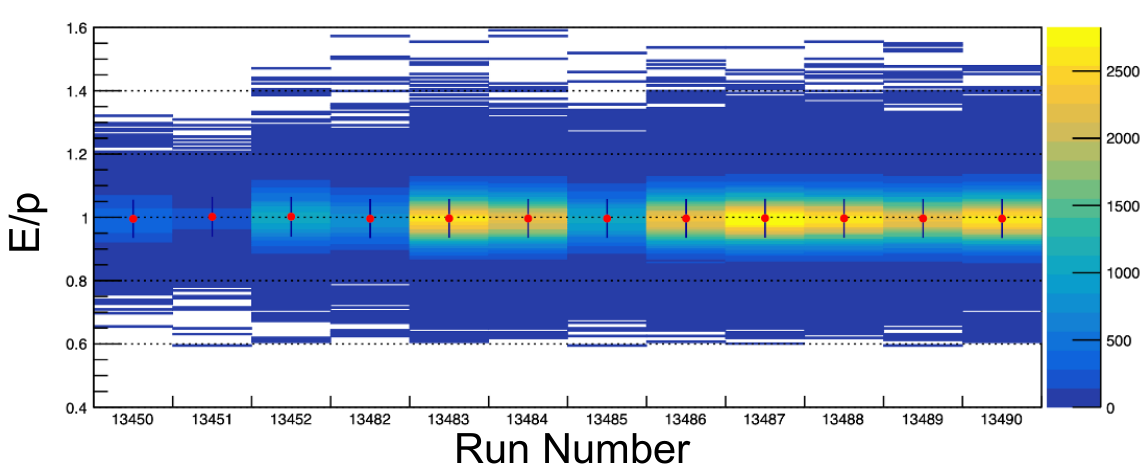}
        \caption{}
        \label{fig:sub-deviation_split1}
    \end{subfigure}
    \hfill
    \begin{subfigure}{0.5\textwidth}
        \centering
        \includegraphics[width= 0.88\textwidth]{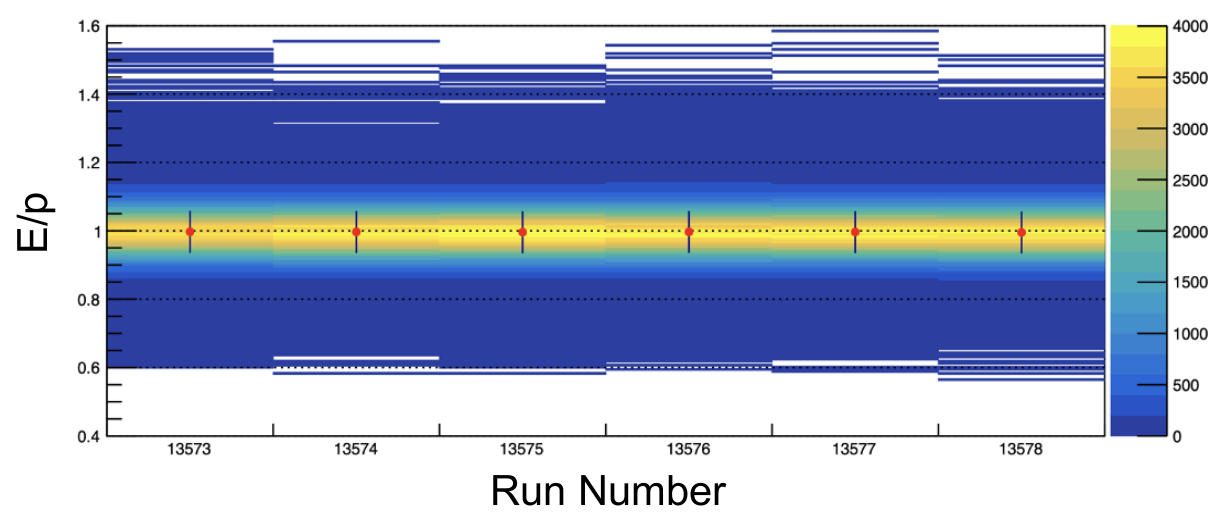}
        \caption{}
        \label{fig:sub-deviation_split2}
    \end{subfigure}
    \caption{The $E/p$ ratio as a function of run number for a selected data set from E12-09-019. The red points represent the mean of the Gaussian fit to each run’s $E/p$ peak. A clear shift in the $E/p$ peak position is observed between runs 13490 and 13573, where the $E/p$ values on either side of this shift are not well aligned at unity (a). When the data are separated into runs taken before (b) and after (c) the shift and calibrated independently, the bias is removed and the $E/p$ peaks for both subsets align at unity as desired.}
    \label{fig:EovP_deviation_issue}
\end{figure*}

The gain coefficients were calibrated separately at each kinematic and magnetic field setting, accounting for the different HV settings used to optimize the trigger. Electrons were distinguished from pions using the PS energy threshold cut discussed in Sec.~\ref{sec:calorimeter}. For some settings, the PS energy cut was augmented by the gas Cherenkov (GRINCH) signals (see Sec.~\ref{sec:introduction}) to improve pion rejection. When possible, the analysis was restricted to elastic scattering events from hydrogen as this gives the cleanest possible selection of electrons with the most accurate momentum reconstruction~\cite{Datta:2024vwq}. Elastic events were selected using a cut around the elastic peak in the photon-nucleon invariant mass spectrum at $W^2 = M_p^2$, where $M_p$ is the proton mass, and optionally also requiring that the corresponding proton be detected in HCal close to its expected position (based on two-body kinematics) with good coincidence timing. 

For settings lacking sufficient elastic data, quasi-elastic and even inelastic events from hydrogen, deuterium and/or $^3$He targets were used, with more aggressive track quality and particle identification cuts to reject false tracks and pions, as well as loose cuts on track momentum and other kinematic quantities to obtain the best possible calibration from the available data. In all cases, an active-area cut was applied to exclude events for which the BBCal cluster center was on the outer edges of the SH layer, ensuring that the clusters used to calibrate BBCal energy were entirely contained within the calorimeter. 


The gain coefficients were determined by minimizing the following $\chi^2$ function:

\begin{equation}
    \chi^2=\sum_{i=1}^N\big( p^i-E^i \big)^2
\end{equation}

\noindent where $N$ is the number of 
events selected for the calibration, $p^i$ is the reconstructed track momentum, and $E^i$ is the total energy of the best BBCal cluster consisting of $M$ PS and SH blocks as defined by:

\begin{equation}
    E^i = \sum_{j=0}^Mc_jA_j^i
\end{equation}

\noindent where, for the $i^{\text{th}}$ event, $c_j$ is the ADC gain coefficient of the $j^{\text{th}}$ block of the cluster, and $A_j^i$ is the block's ADC pulse integral. This $\chi^2$ function was then minimized with respect to the gain coefficients for each PS and SH block:

\begin{equation} \label{eqn:chisq}
    \frac{\partial\chi^2}{\partial c_j}= \sum_{i=1}^N\Big(  A_j^i-\sum_{k=0}^M\frac{A_j^iA_k^i}{p^i}c_k^i \Big) = 0
\end{equation}

Eq.~\ref{eqn:chisq} 
defines 241 simultaneous linear equations 
for the gain coefficients of the 189 SH blocks and 52 PS blocks. This system of equations was solved using standard matrix inversion libraries to obtain the best-fit gain coefficients for all BBCal modules. In some cases, the electron tracks selected for the calibration did not populate the full acceptance of BBCal. To avoid problems with matrix inversion, channels with insufficient statistics and/or small diagonal matrix elements (meaning few hits with sufficiently large signals for an accurate calibration) had their off-diagonal matrix elements forced to zero and their gain coefficients fixed to one, essentially removing them from the calibration. 





After initial energy calibrations were complete, the plots of 
$E/p$ were analyzed run-by-run to check for any possible deviations. Various small deviations were observed throughout individual kinematic configurations, likely due to slight changes in experimental conditions. 
In order to account for these deviations, subgroups of runs that had a significant shift compared to others in the same configuration were calibrated separately.
As can be seen in Fig.~\ref{fig:EovP_deviation_issue}, once the subgroups were separated and recalibrated, they each showed $E/p$ 
peaks centered at one.

\subsubsection{Timing} \label{TimingSec}

\begin{figure*}
    \centering 
    \includegraphics[width=0.95\textwidth]{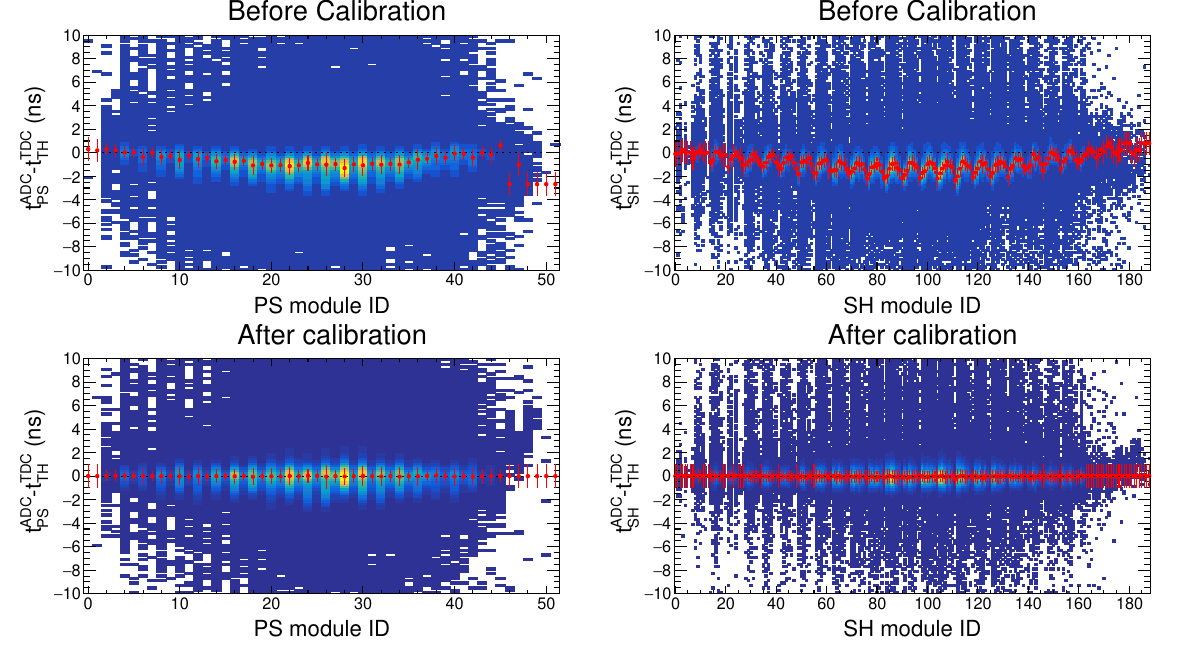}
    \caption{Time difference between TDC time in the scintillator hodoscope and ADC time in the PS (left) and SH (right) versus block number, before (top) and after (bottom) timing alignment. The red markers show the mean of a Gaussian fit to the peak in each channel's time difference distribution. The error bars in all four plots represent the standard deviation of the Gaussian fit, which serves as a measure of the resolution. In the bottom row, channels with insufficient statistics were arbitrarily assigned a mean of zero and a standard deviation of 1~ns for plotting purposes. See text for details.}
    \label{fig:timealign}
\end{figure*}


The leading-edge (LE) time of each BBCal signal pulse was reconstructed directly from the fADC waveform samples using a simple linear interpolation between the two consecutive samples occurring before and after the voltage reaches half of its maximum value on the rising edge of the pulse. For signals well above the software threshold for pulse-finding in the waveform, this algorithm is roughly equivalent to a constant-fraction discrimination, which minimizes time-walk effects in the reconstruction.

Slight non-uniformities in the signal processing circuitry, including but not limited to cable length variations, introduced timing offsets into the PMTs relative to the expected signal arrival times at each BBCal block. Moreover, the sensitivity of the BBCal PMTs to the fringe fields of the SBS magnet required separate timing calibrations for different magnetic field settings.


BBCal ADC times were aligned to the BBS timing hodoscope (see Sec.~\ref{sec:introduction} and Fig.~\ref{fig:bigbitearm}). The typical intrinsic hodoscope resolution after final calibrations was 200-300~ps throughout the SBS program, depending on experimental conditions. However, its effective resolution for time-of-flight measurement is degraded for higher-energy electrons due to the larger spread in both space and time of electromagnetic shower secondaries generated in the PS that contribute to the hodoscope signals. At the same time, the BBCal timing resolution improves as the electron energy increases. At the highest electron energies (approaching 4~GeV) measured by BBS during the SBS program, the effective SH timing resolution equals or even exceeds that of the hodoscope. 
The BBCal-hodoscope alignment was achieved via Gaussian fits to the distributions of the time differences between individual PS and SH ADC channels and the hodoscope mean cluster time, after the latter was corrected for electron time-of-flight variation within the acceptance and aligned to the accelerator RF signal.


As shown in Fig.~\ref{fig:timealign}, after alignment, the residual channel-to-channel variations in the BBCal-hodoscope time differences were small compared to the combined intrinsic resolution of the detectors. In Fig.~\ref{fig:timealign}, some channels in the periphery of the acceptance were not successfully calibrated due to insufficient statistics. For these channels, an offset was assigned based on an average of the nearest successfully calibrated neighboring channels, with somewhat mixed results. Manual, post hoc adjustments were made on some channels with poor data and/or fit quality as needed. For the most part, poorly-calibrated channels were outside the useful acceptance anyway. 
\begin{figure}
    \begin{center}
        \includegraphics[width=0.48\textwidth]{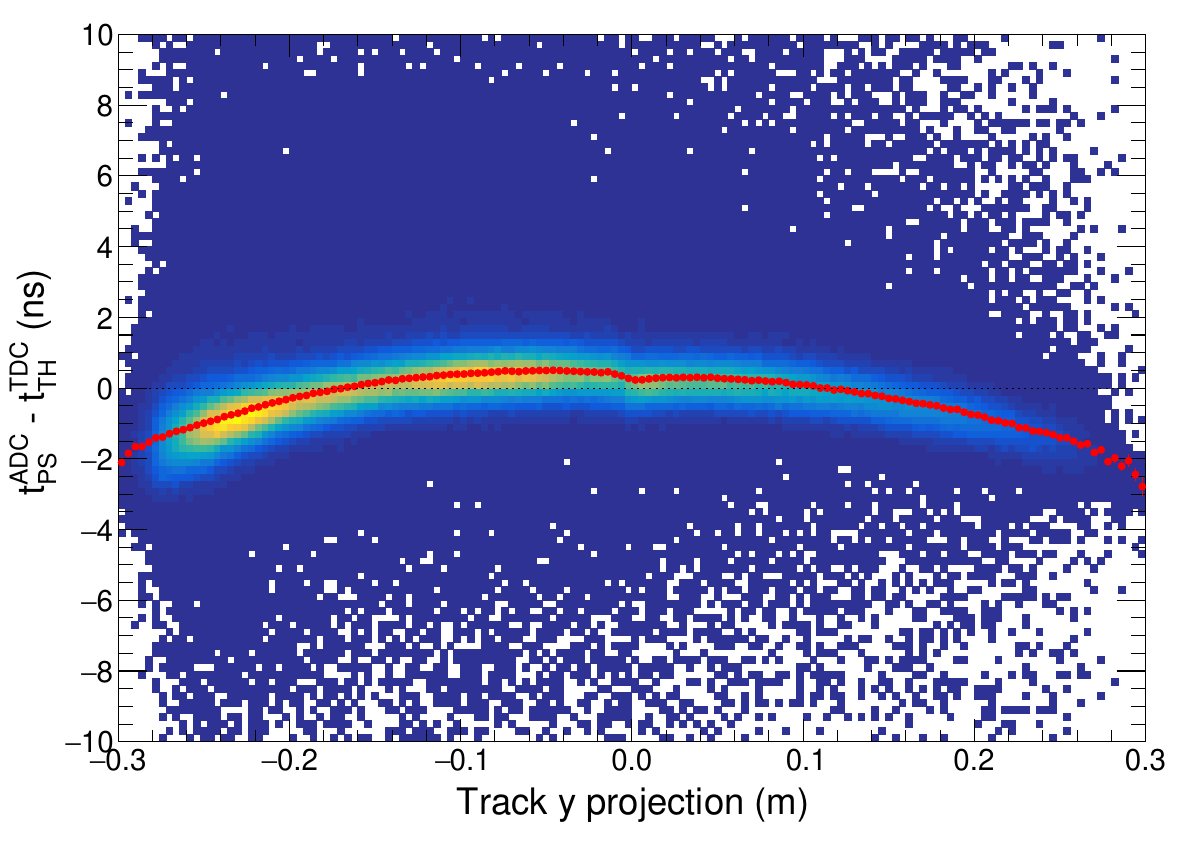}
    \end{center}
    \caption{\label{fig:dtpshodotracky} PS-hodoscope time difference versus the horizontal position of the track projection to the PS. Red markers show the approximate mean of the distribution at each position. Tracks crossing the PS closer to the PMTs give earlier arrival times as naively expected. The kink in the distribution near the center reflects the boundary between the left and right columns of the PS. Note that the direction of the $y$ coordinate in this figure runs from small to large scattering angles; i.e., from left to right as viewed from downstream, when BBS is on beam left.}
\end{figure}

As can be seen in Fig.~\ref{fig:timealign}, the distributions of PS-hodoscope time differences are slightly skewed/asymmetric, even after alignment. This reflects a neglected spurious correlation between the PS ADC time and the horizontal position at which the electron track crosses the PS. Fig.~\ref{fig:dtpshodotracky} shows an example of this correlation from E12-09-019 data. The observed correlation is suggestive of light propagation delay being the main physical mechanism; tracks crossing the PS blocks closer to the PMTs produce earlier signals. The magnitude of the effect (about 2~ns variation over the $\sim$30-cm length of the blocks) is also qualitatively consistent with this interpretation, given the expected effective light propagation speed in LG, accounting for the refractive index and the angular distribution of the Cherenkov radiation.   

\begin{figure}[h!]
    \begin{center}        
        \includegraphics[width=0.48\textwidth]{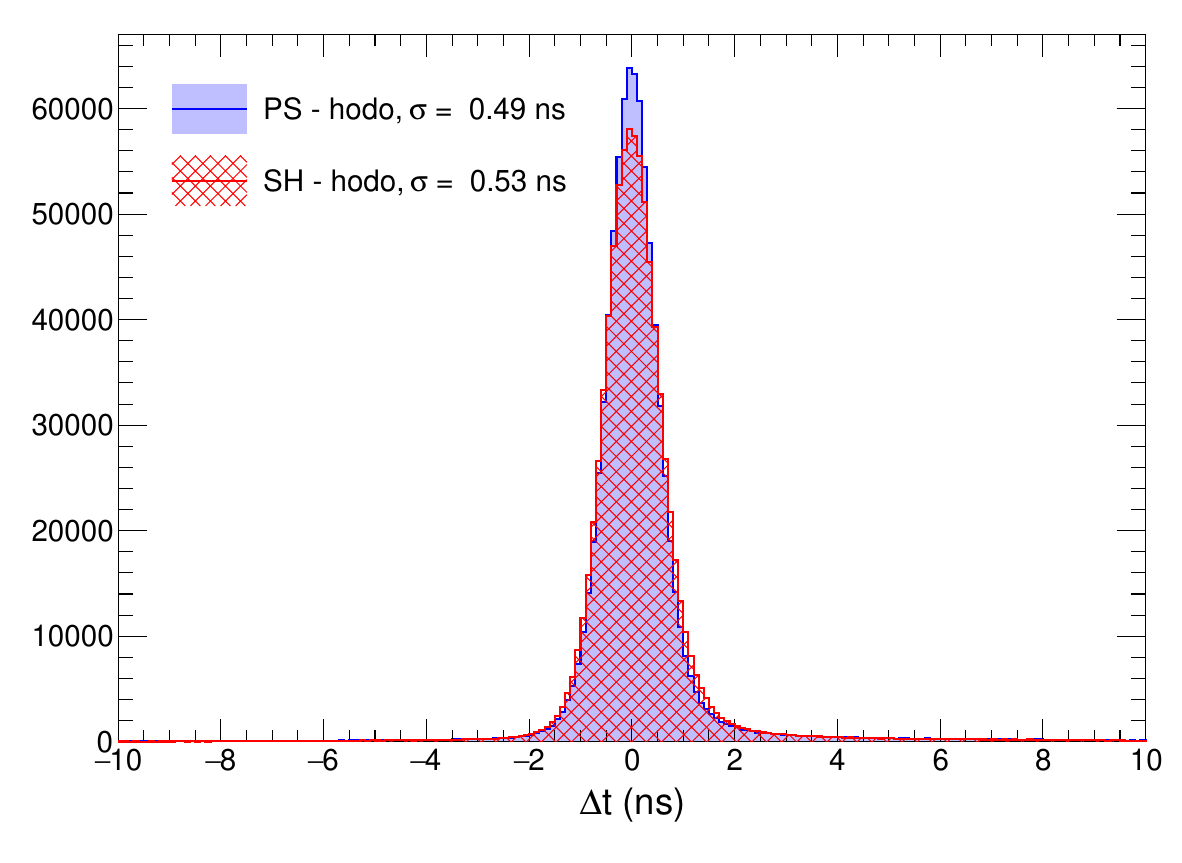}
    \end{center}
    \caption{\label{fig:dtBBCALhodo} PS-hodoscope (blue solid) and SH-hodoscope (red hatched) time difference distributions after final calibrations, with Gaussian fits to estimate the resolution. The PS time has been corrected for the correlation with horizontal position shown in Fig.~\ref{fig:dtpshodotracky}. Assuming 250-ps resolution for the hodoscope, the implied PS (SH) timing resolution is $\sigma_t\approx0.43(0.48)$~ns.}
\end{figure}

One way of estimating the intrinsic time resolution of BBCal is to compare the SH and PS ADC times to the hodoscope after final alignments. Fig.~\ref{fig:dtBBCALhodo} shows a representative example of the time difference distributions between the SH/PS and the hodoscope after final calibrations, for data from E12-09-019. The PS times shown in Fig.~\ref{fig:dtBBCALhodo} were corrected for the correlation shown in Fig.~\ref{fig:dtpshodotracky} before subtracting the hodoscope times. While the widths shown in Fig.~\ref{fig:dtBBCALhodo} are determined by the combined resolution of the hodoscope and BBCal, they are thought to be mostly dominated by the BBCal timing resolution, as the estimated (intrinsic) hodoscope mean-time resolution for this setting was 250~ps. The comparisons above suggest the typical BBCal time resolution is 0.4-0.5~ns.

\begin{figure}
    \centering
    \includegraphics[width= 0.95\linewidth]{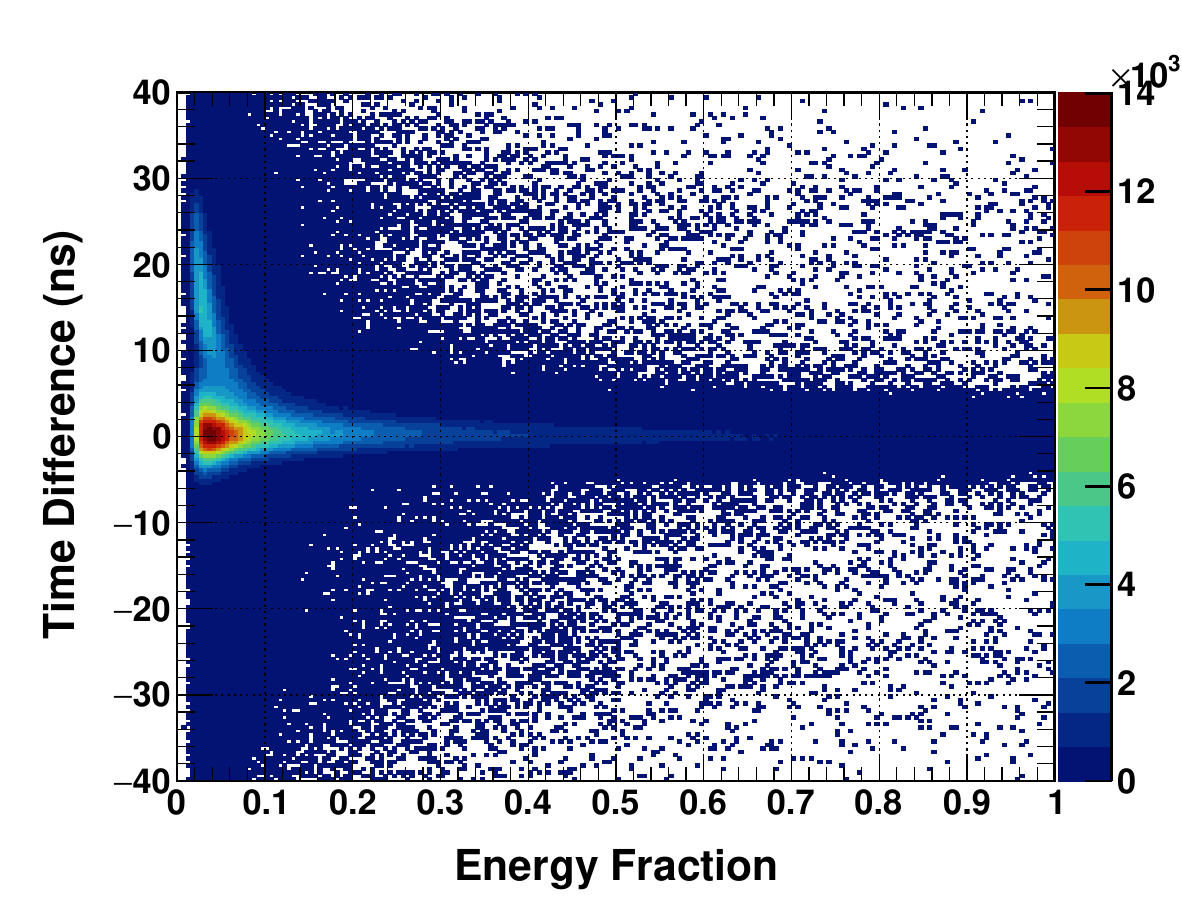}
    \caption{Correlation between the time difference of the secondary and primary blocks in a SH cluster and the ratio of the secondary block’s energy to that of the primary, with low-energy out-of-time hits clearly visible. Figure adapted from \cite{Datta:2024vwq}.}
    \label{fig:energy-ADCdiff}
\end{figure}

Furthermore, double peaking was observed in the time difference distribution between secondary and primary blocks of SH clusters, as can be seen in Fig.~\ref{fig:energy-ADCdiff}. These out-of-time hits were low energy, approximately $1\%$ of the highest energy hit, and occurred when the secondary block was in a separate row from the primary block. Thus, the likely cause of these out-of-time hits is multiple scattering of low-energy events from the main electromagnetic shower. 

The fact that out-of-time hits were only seen in the SH when the cluster spanned over two SH rows supports the idea that the mu-metal between the SH rows increased the probability of multiple scattering. We found that a simple $\pm10$~ns cut on the ADC time difference between primary and secondary blocks in a cluster not only got rid of the out-of-time hits, but also improved the energy resolution across all kinematic configurations.

\section{Simulation}
\label{sec:simulation}

Analysis of SBS experiment data requires a realistic Monte Carlo (MC) simulation  framework to accurately model relevant physics processes and detector effects. The \textit{g4sbs} framework~\cite{G4SBS} comprises event generators for all relevant physics processes and a detailed Geant4~\cite{GEANT4:2002zbu} geometry of the experimental setup—including the targets, spectrometers, beamlines, and magnetic fields. The companion C++ library \textit{libsbsdig}~\cite{SBSDIG} digitizes events simulated by \textit{g4sbs} to emulate signal processing effects, producing pseudo-raw data that is then reconstructed by the same algorithms used to process the real data~\cite{SBSoffline}. Together, this software infrastructure enables direct and meaningful comparisons between simulated and experimental data.

\textit{g4sbs} was used extensively during the proposal, design, planning, execution, and analysis phases of all SBS experiments. Prior to E12-09-019, \textit{g4sbs} played a crucial role in estimating the BBCal trigger rates \cite{BBSRates2025} and thresholds, supporting the use of a single-arm electron trigger for the E12-09-019 measurement --- an important requirement for controlling systematic errors. During data analysis, \textit{g4sbs} is indispensable for the extraction of physics observables from the data~\cite{Datta:2024vwq}.

\begin{figure}[h!]
    \centering
    \includegraphics[width= 0.98\linewidth]{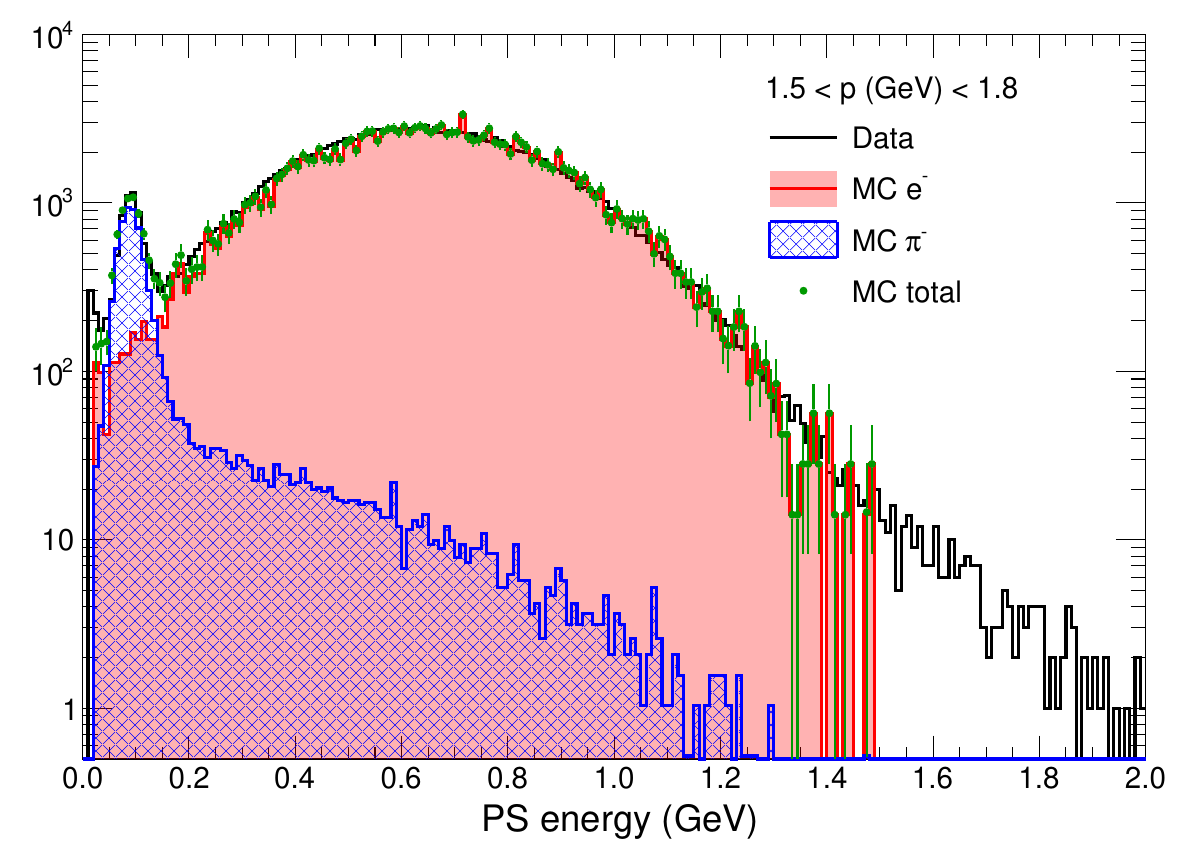}
    \caption{Data-simulation comparison of the PS energy distribution for negatively-charged tracks in the 1.5-1.8 GeV momentum range produced in the scattering of 4~GeV electrons on the 15~cm hydrogen target. The black histogram shows the experimental data. The red shaded (blue hatched) histogram shows the simulated PS response to electrons (charged pions). The green dotted histogram shows the sum of simulated electron and pion distributions. See text for details.}
    \label{fig:PSenergySIM}
\end{figure}

\begin{figure*}[t]
    \begin{center}
        \includegraphics[width=0.95\textwidth]{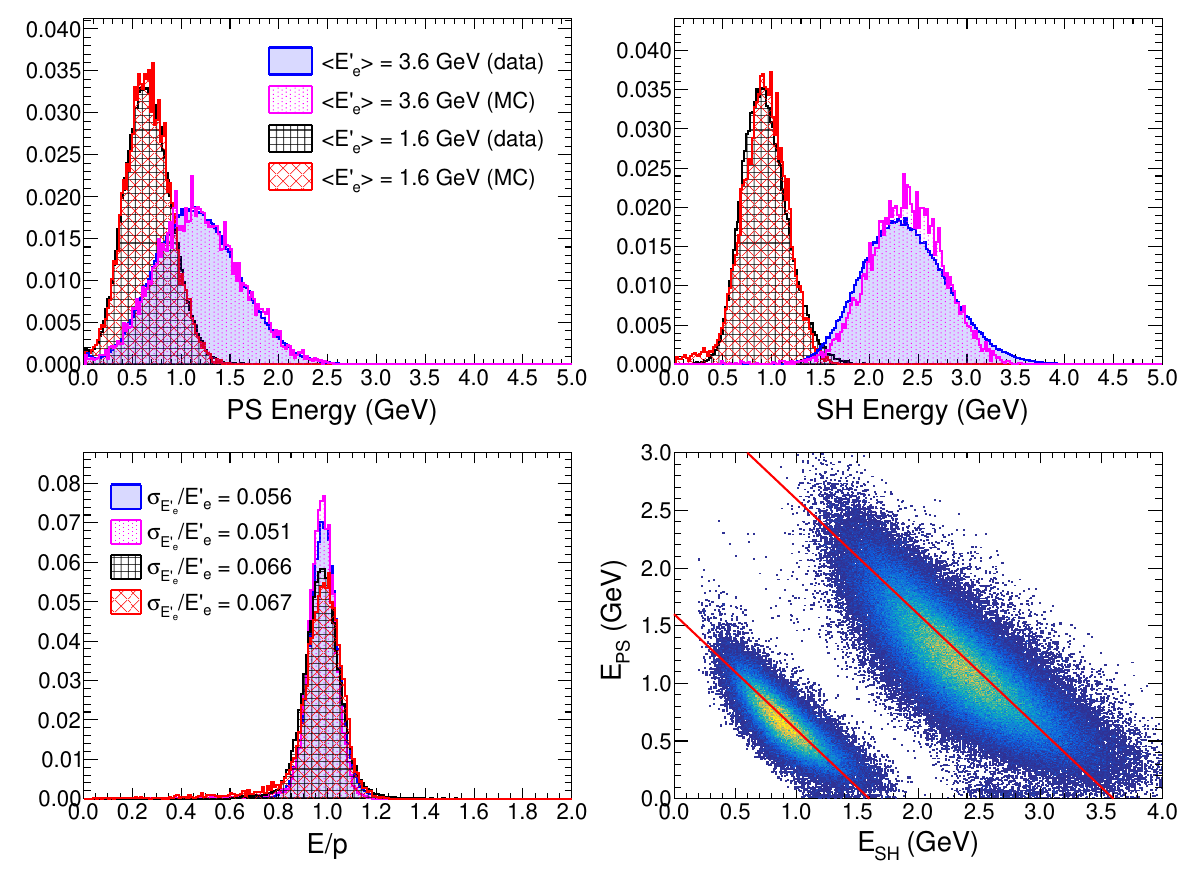}
    \end{center}
    \caption{\label{fig:BBCALsimdatacomp} Comparison of observed (``data") and simulated (``MC") BBCal responses to elastically scattered electrons from hydrogen, at average energies of 1.6 and 3.6 GeV incident on BBCal. The top left (top right) plot shows the energy distribution in the PS (SH), with electron energies and data/MC correspondences as indicated in the legend, with the $y$ axis normalized to the fraction of events per 0.02~GeV interval. The bottom left plot shows the BBCal energy resolution, as measured by the ratio $E/p$ of the reconstructed energy $E$ in BBCal to the momentum $p$ reconstructed from tracking results (see Sec.~\ref{subsec:beamcal}), with the $y$ axis normalized to the fraction of events in an $E/p$ interval of 0.01. The quoted fractional resolutions indicated in the legend of the bottom left plot are the standard deviations of Gaussian fits to the distributions in the range $0.9 < E/p < 1.1$, with the same data/MC and energy correspondences as indicated in the legend of the top left plot. The bottom right plot shows the correlation between PS and SH energies from the real data for both settings. The color scale is proportional to the number of events, but the relative normalization between 1.6 GeV and 3.6 GeV distributions is arbitrary, with roughly similar total numbers of events included at each energy. The red diagonal lines are plotted at $E_{PS}+E_{SH} = 1.6$ and $3.6$ GeV, and are intended to guide the eye. See text for details.}
\end{figure*}

Fig.~\ref{fig:PSenergySIM} shows an example PS energy distribution from the E12-09-019 experiment for events producing a negatively charged track in BBS from the 15-cm LH$_2$ target in the 1.5-1.8 GeV/c momentum range, which coincides with the energy range of elastically scattered electrons for $E_{beam} = 4.0$ GeV, $Q^2 = 4.5$ (GeV/c)$^2$ (see Tab.~\ref{tab:CF}). The black histogram shows the experimental data, with a broad peak from high-energy electrons and a sharp peak at low energy from MIPs (mainly $\pi^-$). To describe and interpret these features, elastically scattered electrons (red shaded histogram) and negatively charged pions in the relevant momentum range (blue hatched histogram) were generated separately and tracked through the \textit{g4sbs} model of the BBS. To facilitate direct comparison to the data, the simulated events were then processed by the aforementioned digitization and reconstruction libraries. The green dotted histogram, representing the sum of simulated electrons and pions, shows good agreement with the data, indicating accurate modeling of the detector response.

Note that while the real data shown in Fig.~\ref{fig:PSenergySIM} contain a mix of elastically and inelastically scattered electrons and pions in the 1.5-1.8 GeV momentum range, the simulated electrons and pions are not identically distributed in terms of kinematics. In particular, the simulated electron distribution only includes elastic scattering events from hydrogen, while the simulated pions are generated uniformly in momentum and solid angle within the BBS acceptance. Since the relative normalizations between the real PS energy spectrum and the simulated electron and pion distributions shown in Fig.~\ref{fig:PSenergySIM} are arbitrary, the simulated pion and electron distributions were normalized separately such that their sum matches the data. 

In general, the PS response to ultra-relativistic electrons, determined by the  rates of Bremsstrahlung and secondary $e^+/e^-$ pair production, exhibits a significant energy dependence. For charged pions, the dominant MIP-like peak at around 89 MeV is roughly energy-independent, but the high-energy tail, corresponding to the small fraction of pions undergoing hadronic collisions in the PS, is not. As such, one should not generally expect \textit{perfect} agreement between simulated and observed BBCal responses to electrons and/or charged pions unless the simulated and measured particle ID and momentum distributions are identical. However, the restriction of the analysis to a relatively narrow range of track momenta leads to very good qualitative and quantitative agreement in this example. In the real data analysis, the GRINCH (see Fig.~\ref{fig:bigbitearm} and Sec.~\ref{sec:introduction}) and the constraints from two-body elastic/quasi-elastic scattering kinematics provide strong validation and cross-checks of the (simulation-informed) $\pi/e$ separation based on the preshower response.

Fig.~\ref{fig:BBCALsimdatacomp} compares observed and simulated BBCal responses to elastically scattered electrons from hydrogen, for the kinematics with the lowest and highest scattered electron energies from Tab.~\ref{tab:CF}, illustrating the accurate qualitative and quantitative description of the BBCal response to electrons. As described in~\cite{Datta:2024vwq} and Sec.~\ref{subsec:beamcal}, elastic $ep$ scattering events are selected from the real hydrogen target data using the virtual photon-nucleon invariant mass $W$ and the angular and timing correlations between the electron and proton detected in BBS and SBS, respectively. The simulation reproduces the observed PS and SH energy distributions and the total shower energy resolution remarkably well at both energies, considering the complexity of both the simulation $\rightarrow$ digitization $\rightarrow$ reconstruction software chain and the real detector hardware and its associated signal processing electronics.

According to the simulation, ultra-relativistic electrons that do not suffer large radiative energy losses before reaching BBCal deposit an average of about 12\% of their energy (with a standard deviation of approximately 3\%) in materials other than the BBCal lead-glass, including target and scattering chamber exit window materials, air, the other BBS detectors, and passive shielding and support structures upstream of BBCal, with the latter giving the dominant contribution. In particular, the iron shielding plates enclosing the PS (1/2-inch total thickness or $\approx 70\%$ of a radiation length) account for most of the non-BBCal energy losses for high-energy electrons. This ``undetected" energy fraction of roughly 12\% also includes any ``leakage" of the SH energy outside of BBCal itself and is approximately independent of electron energy within the 1-4 GeV range relevant to the SBS program. Moreover, it is worth noting that a non-trivial fraction (typically 2-3\% with significant random fluctuations) of the energy from the electromagnetic cascade that starts in the PS is absorbed by the timing hodoscope, which lies between the PS and the SH\footnote{The timing hodoscope is only equipped with timing readout, lacking amplitude readout. As such, it cannot directly contribute to the electron energy measurement.}. 

In the real data analysis, the BBCal gain coefficients (see Sec.~\ref{subsec:beamcal}) are calibrated to the reconstructed electron momentum from tracking, assuming the electrons deposit all their energy in BBCal. As such, the ``undetected" energy, which is not directly measurable on an event-by-event basis, is effectively absorbed into the BBCal energy calibration, and its random fluctuations contribute to the energy resolution (and the generally very small non-linearities in the relationship between energy deposition and signal amplitude). The data-MC comparisons shown in Fig.~\ref{fig:BBCALsimdatacomp} are based on the effective calibrated energy reconstruction in BBCal. For the simulation, a single effective gain factor applied to the simulated, digitized, and reconstructed energy in the PS and SH that accounts for this ``missing" energy suffices for a direct comparison to the calibrated energy reconstruction from the real data. 

Several features of the comparisons shown in Fig.~\ref{fig:BBCALsimdatacomp} deserve explicit mention. First, 3.6-GeV electrons deposit, on average, nearly twice as much energy in the PS as 1.6-GeV electrons, consistent with the rising rates of radiative energy losses in a given thickness of lead-glass as the incident electron energy increases. Secondly, while the simulation achieves a nearly perfect description of the PS responses at both energies and of the shower response at 1.6 GeV, the SH energy distribution for 3.6-GeV electrons is noticeably wider and more asymmetric in the real data than in the simulation, leading to a slight mismatch in energy resolution, with the simulation predicting a slightly better resolution of 5.1\% than the 5.6\% \footnote{The slight differences in energy resolution between the data shown in Fig.~\ref{fig:BBCALsimdatacomp} and quoted in Tab.~\ref{tab:enRes} result from slight differences in data selections, histogram binnings, and Gaussian fits and are neither significant nor meaningful.} seen in the (calibrated) data. This is thought to be partially attributable to calibration uncertainties that affect the real data but not the simulation and to the effect of the SBS fringe field on the SH PMT performance, which is known to be significant for this kinematic setting, in which the BBS and SBS were very close together.


\section{Performance}

\subsection{Energy Resolution}

\begin{figure}[h!]
    \centering
    \includegraphics[width=\linewidth]{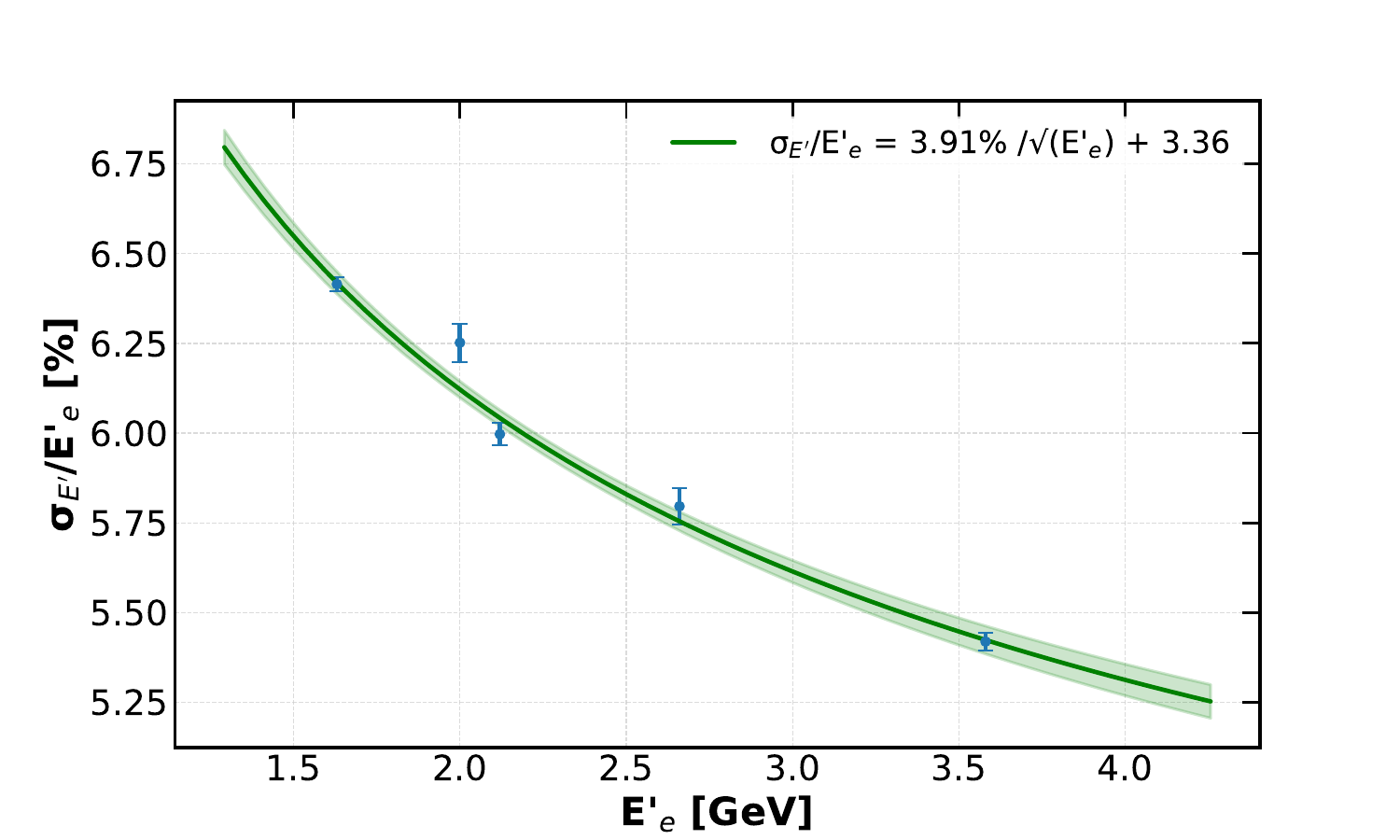}
    \caption{Energy resolution versus central elastic electron energy for kinematic points in E12-09-019. There is a 1-$\sigma$ error band on the fit.}
    \label{fig:corr_perf_plot}
\end{figure}

The energy resolution of BBCal is measured by the $E/p$ distribution, e.g., Fig.~\ref{fig:calib-compare}. Because the momentum resolution of BBS is $1-1.5\%$, the width of the $E/p$ distribution is dominated by the BBCal energy resolution. We define $\sigma_{E'}/E'_e$, the standard deviation of the Gaussian fit to the $E/p$ peak, as the energy resolution of the calorimeter. Fig.~\ref{fig:corr_perf_plot} shows the energy resolution as a function of the central elastic electron energy, with the numerical values given in Tab.~\ref{tab:enRes}. The data have been fitted using:

\begin{equation}\label{perfEqn}
    \frac{\sigma_{E'}}{E'_e}= \frac{(3.9\pm0.1)} {\sqrt{E'_e}}\% +(3.4\pm0.2)\%    
\end{equation}

\noindent which can be compared to the similar fit parameterization found in Sec.~3 of \cite{AVAKIAN199869}.

\begin{table}[]
    \caption{BBCal energy resolution values for different kinematic points measured during E12-09-019.   $E_{\text{beam}}$ is the electron beam energy, $E'_{e}$ is the central scattered electron energy, $Q^2$ is the central $Q^2$, and $\sigma_{E'}/E'_e$ is the measured BBCal energy resolution. Note that the quoted ``error" on the resolution is purely statistical and should not be taken as a measure of uncertainty of the resolution.}
    \label{tab:enRes}
    \centering 
    \begin{tabular}{cccc}
        \hline\hline\vspace{-1em} \\ 
        $E_{\text{beam}}$ (GeV) & $E'_{e}$ (GeV) & $Q^2$ (GeV/c)$^2$ & $\frac{\sigma_{E'}}{E'_e}$ (\%) \vspace{0.2em} \\ \hline \vspace{-1.1em} \\
        4.0 & 1.6 & 4.5 & 6.42 $\pm$ 0.02 \\
        6.0 & 2.0 & 7.4 & 6.25 $\pm$ 0.05 \\
        3.7 & 2.1 & 3.0 & 6.00 $\pm$ 0.03 \\
        7.9 & 2.7 & 9.9 & 5.80 $\pm$ 0.05 \\
        9.9 & 2.7 & 13.6 &  6.43 $\pm$ 0.17 \\
        6.0 & 3.6 & 4.5 & 5.42 $\pm$ 0.03 \\
        \hline\hline
    \end{tabular}
\end{table}

Note that the BBCal energy resolution quoted in Tab.~\ref{tab:enRes} for beam energy $E_e=9.9$~GeV and $Q^2=13.6$~GeV$^2$, with central elastically scattered electron energy $E'_e=2.7$~GeV, is significantly worse than that obtained at $E_e=7.9$~GeV and $Q^2=9.9$~GeV$^2$ for the same $E'_e$ and would lie significantly above the best-fit curve through the other data points in Fig.~\ref{fig:corr_perf_plot}. This is attributable to several factors. First, the calibration of this dataset was less accurate due to low elastic $ep$ statistics and poorer electron identification due to very unfavorable $\pi/e$ ratios prevailing at such large $Q^2$. Second, and more importantly, the SBS was positioned at a forward angle of $13.3^\circ$ for this kinematic setting, putting it very close to the BBS, and it was operated at its maximum field setting, leading to a large fringe field at the location of the BBCal PMTs, which further degraded the PMT performance and likely impacted the energy resolution. As such, this point is not included in the fit shown in Fig.~\ref{fig:corr_perf_plot} and quoted in Eqn.~\ref{perfEqn}. 


The first term in Eqn.~\ref{perfEqn} is a stochastic term that accounts for shower intrinsic fluctuations~\cite{RevModPhys.75.1243}. The second term is a constant term that accounts for inhomogeneities in the detector that may arise from calibrations or geometry. The random fluctuations of the "undetected" energy deposited outside of BBCal (see sec.~\ref{sec:simulation}) also contribute. This constant term also gives an idea of the best resolution we can possibly achieve. As the electron energy increases, we reach a lower limit on our energy resolution of approximately $3.4\%$. There is often a third term in standard energy resolution fits that accounts for noise in the signal. This noise term was excluded in Eqn.~\ref{perfEqn} due to the small energy range in which BBCal was used in SBS and the generally high signal to noise ratios in BBCal. The weighted average of all the energy resolution values shown in Fig.~\ref{fig:corr_perf_plot} is approximately $6.2\%$.

\subsection{Position Resolution}

The position resolution of BBCal can be measured using the highly precise tracking system which sits upstream of the PS layer. The design of the PS layer naturally leads to poor position resolution in the horizontal direction. Due to the high background rates in the GEMs, both the spatial and energy resolution of BBCal played an important role in defining the smallest possible search region for track-finding in the GEMs, with the spatial resolution of the shower playing the dominant role. 

The tracking system in E12-09-019 has a position resolution of approximately $70~\mu$m and an angular resolution significantly better than 1~mrad~\cite{GNANVO201577, ALTUNBAS2002177}. As such, the position of the track at BBCal is known with very high precision. Comparing the SH coordinates measured by BBCal to those measured by the tracking system gives an accurate measure of the BBCal spatial resolution, with negligible contribution from the tracking resolution. As noted in Sec.~\ref{sec:clus}, the position of a cluster in the SH is defined as the energy-weighted centroid of the blocks included in the cluster.


\begin{figure}[h!]
    \centering
    \includegraphics[width= 0.48\textwidth]{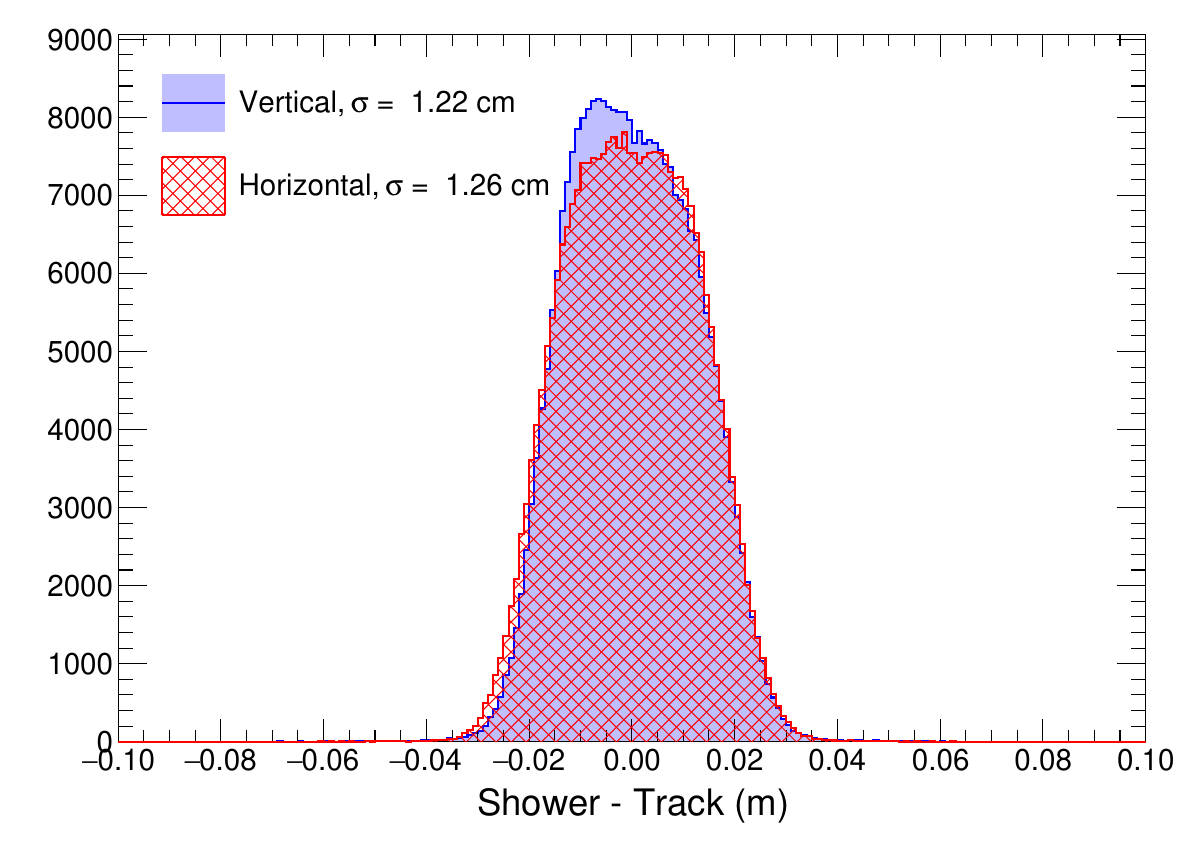}
    \caption{Difference between reconstructed SH coordinate and projected track coordinate along the dispersive (vertical) and non-dispersive (horizontal) directions, for elastically scattered electrons with an average energy of 3.6 GeV. The quoted resolution is the standard deviation of the distribution. See text for details.}
    \label{fig:SHposRes}
\end{figure}


Fig.~\ref{fig:SHposRes} shows a representative example of SH spatial resolution, as measured by comparison to the projected track position at the SH layer. The non-Gaussian distribution of this difference owes to the relatively large size of the SH blocks compared to the Moli\`ere radius of lead-glass, which limits the spatial resolution that can be obtained from BBCal. The slight asymmetry of the distributions along both directions results from nonuniformities in the (acceptance-averaged) distribution of events within the primary block. The resulting spatial resolution is 1.2-1.4~cm, with a slight improvement at higher electron energies. While the shower spatial resolution could be improved slightly with more sophisticated position reconstruction methods, the resolution obtained using the standard energy-weighted average block position in the SH cluster proved more than sufficient for the tracking needs of the BBS. Moreover, the resolution of the region-of-interest calculation for the upstream GEM layers is not solely determined by the shower spatial resolution, but also has important contributions from the BBCal energy resolution and the target thickness for the vertical and horizontal projections, respectively.

\subsection{Timing Resolution}

In Sec.~\ref{TimingSec}, the time difference between BBCal and the BBS timing hodoscope was used as a proxy for the BBCal timing resolution, suggesting a resolution of 0.4-0.5~ns. Other ways to estimate the BBCal time resolution include comparing the SH and PS timing, or examining the time differences between neighboring blocks within a given BBCal cluster. For the scattered electrons in the SBS experiments, it is naively expected that two neighboring blocks within a cluster should fire at approximately the same time, so the difference between the ADC times of these two blocks should be dominated by the timing resolution of the calorimeter. Thus, the time differences between blocks within a cluster, or between SH and PS blocks, serve as a useful cross-check of the estimates based on comparisons to the hodoscope presented in Sec.~\ref{TimingSec}. 

\begin{figure}
    \begin{center}
        \includegraphics[width=0.48\textwidth]{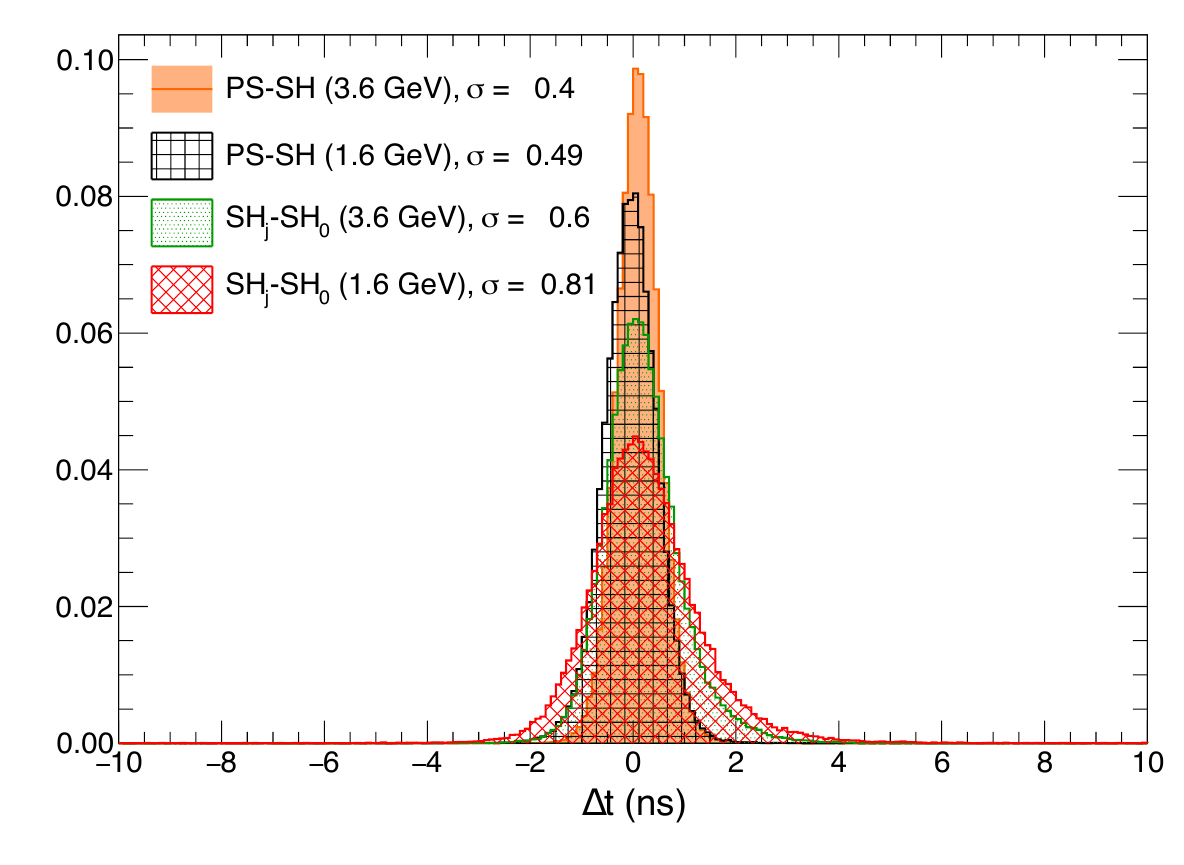}
    \end{center}
    \caption{\label{fig:dtPSSH} Time differences between highest-energy PS and SH blocks (``PS-SH" in the legend) and between secondary and primary blocks in the SH cluster (``SH$_{\text{j}}$-SH$_0$" in the legend), for elastically scattered electrons with average energies of 3.6 and 1.6~GeV, illustrating the energy dependence of BBCal effective timing resolution. The quoted resolutions are the standard deviations (in ns) of Gaussian fits to the respective histograms. The histograms are normalized such that the $y$ axis scale represents the fraction of events per 0.01-ns interval. See text for details. }
\end{figure}
Fig.~\ref{fig:dtPSSH} shows four time difference distributions based on comparisons of different BBCal signals over a range of electron energies. Events selected for the analysis shown in Fig.~\ref{fig:dtPSSH} were elastically scattered electrons from hydrogen, with average energies of 3.6 and 1.6~GeV. The time difference between the highest-energy blocks in the PS and SH clusters shows a width of approximately 0.4 (0.5)~ns at 3.6 (1.6)~GeV, consistent with the comparisons to the timing hodoscope shown in Sec.~\ref{TimingSec} (which used 1.6-GeV electrons). The time difference between secondary blocks and the primary block in the SH cluster shows a larger width of about 0.6 (0.8)~ns at 3.6 (1.6)~GeV. This is attributable to the generally poorer timing resolution for small signals in the periphery of SH clusters, as compared to the primary block, which tends to give the best timing resolution. We can also make such comparisons for primary and secondary blocks in the PS. This gives similar results as for the SH; however, the majority of PS clusters consist of exactly one block (see Fig.~\ref{fig:clusMult}), except for events near the boundaries between rows and columns. 

 At 1.6 GeV the total energy deposit in BBCal is divided fairly evenly between the SH and the PS (see Fig.~\ref{fig:BBCALsimdatacomp}). For both settings, the electrons were required to deposit at least 0.2~GeV in both the SH and PS, and to have an $E/p$ ratio between $0.8<E/p<1.2$ (see Figs.~\ref{fig:calib-compare} and \ref{fig:BBCALsimdatacomp}). The energy deposition in the secondary blocks was also required to be at least 20\% of the primary block's energy deposit and only secondary blocks immediately adjacent to the primary block were considered. No additional requirements were imposed on either the primary or secondary blocks' energies. As such, many hits with relatively low energies are included in this comparison, explaining its generally larger width, consistent with the energy dependence of time differences shown in Fig.~\ref{fig:energy-ADCdiff}. 


At higher secondary block energies, the width of the time difference distribution between primary and secondary SH blocks approaches that of the SH-PS, SH-hodoscope, and PS-hodoscope comparisons. Apart from the significantly delayed out-of-time hits in the SH clusters, as seen in Fig.~\ref{fig:energy-ADCdiff}, there is no evidence for a significant time-walk effect at low energies with the standard reconstruction algorithm for leading-edge times; instead, it seems the time resolution for individual SH blocks is simply degraded for low-energy hits with smaller signal/noise ratios. In practice, the overall effective timing resolution for BBCal, taking the highest-energy blocks in both the SH and PS, was 0.4-0.5~ns, generally improving at higher electron energies.

\begin{figure}
    \begin{center}
        \includegraphics[width=0.48\textwidth]{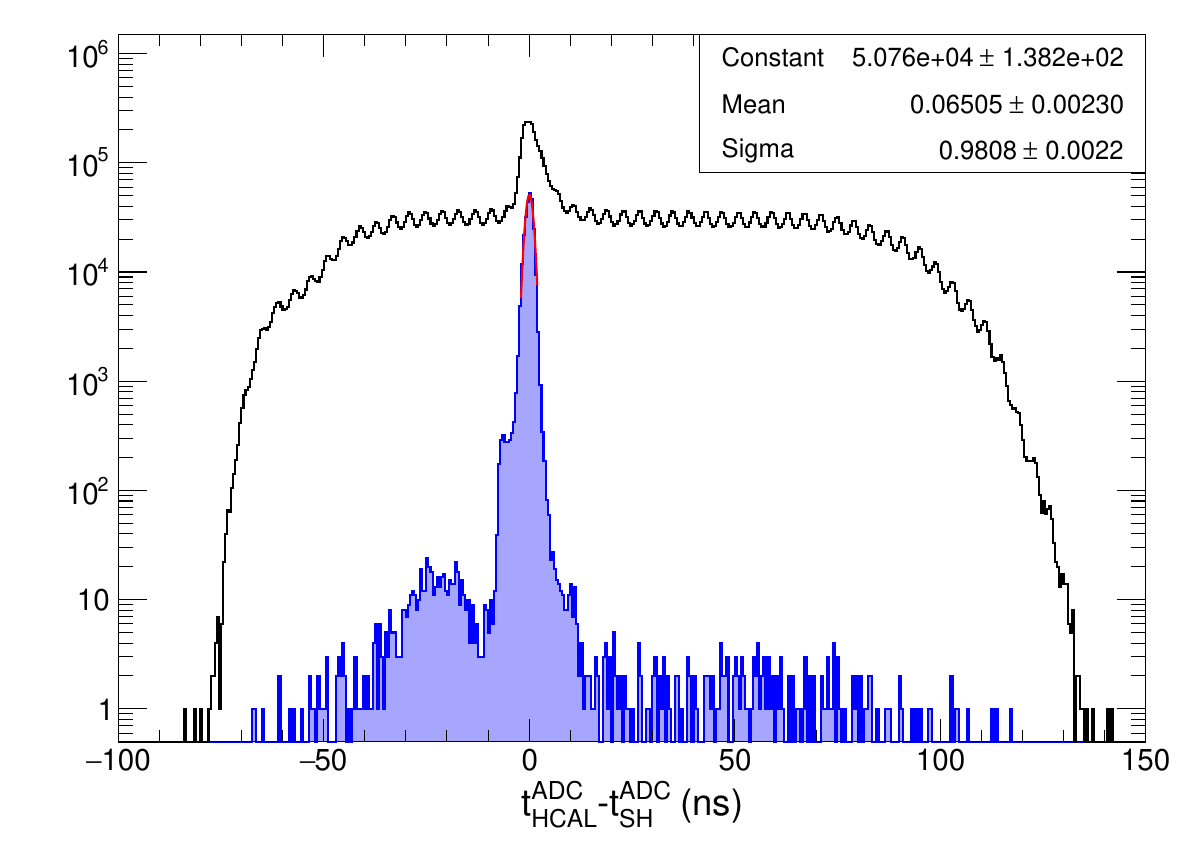}
    \end{center}
    \caption{\label{fig:coin} Difference between HCal and BBCal ADC times. The black histogram shows all events with a good electron in BBS and any particle detected in HCal, with real coincidence peak and accidental background showing the 4-ns beam bunch spacing of CEBAF used during E12-09-019. The blue shaded histogram with Gaussian fit shows the same distribution for elastic $ep \rightarrow ep$ scattering events, selected as described in Sec.~\ref{subsec:beamcal} and Sec.\ref{sec:simulation} (and Ref.~\cite{Datta:2024vwq}), illustrating the strong suppression of accidental backgrounds and the coincidence time resolution of approximately 1~ns, dominated by HCal.}
\end{figure}
During the SBS experiments, the coincidence time between when BBCal measured an event and when HCal (see Sec.~\ref{sec:introduction}) measured an event was used to define in-time events of interest. While some of the experiments used a single-arm trigger based on BBCal alone, others used a coincidence between BBCal and HCal. While this paper will not go into detail regarding the hadron calorimeter, the resolution of this coincidence time was extremely important for the suppression of accidental and inelastic backgrounds. The typical resolution of coincidence time was approximately 0.8-1~ns, dominated by HCal. Fig.~\ref{fig:coin} shows a representative example of the HCal-BBCal coincidence time distribution, before and after elastic $ep$ event selection, at $Q^2 = 4.5$ GeV$^2$, $E'_e = 3.6$ GeV. For the elastic events, the non-Gaussian ``shoulders" of the real coincidence peak and the asymmetric distribution of accidental coincidences surviving the cuts reflect the effects of small numbers of miscalibrated SH and/or HCAL channels and/or hits with bad timing reconstruction, and these events are rejected from the analysis by the coincidence time cut.  

For the setting shown in Fig.~\ref{fig:coin}, accidental coincidences are suppressed by more than three orders of magnitude (relative to the inclusive electron sample) for events identified as elastic $ep \rightarrow ep$ based on the measured scattering angles and/or energies of both detected particles, making the coincidence time cut essentially redundant in terms of background rejection power. On the other hand, at higher $Q^2$ values, the cross sections for the quasi-elastic $A(e,e'N)$ scattering processes of interest for neutron form factor measurements are much more strongly suppressed relative to inelastic background processes as compared to the Fig.~\ref{fig:coin} example. Moreover, the two-body kinematic correlations between the outgoing electron and nucleon are smeared by the Fermi motion of the bound nucleons in the deuterium and $^3$He targets used in E12-09-019 and E12-09-016 respectively. For these measurements, the HCal-BBCal coincidence time played an extremely important role in event selection and background rejection. The combined resolution of BBCal and HCal was sufficient to resolve the 4-ns beam bunch spacing in the accidental coincidence distribution after both detectors' timing offsets were properly aligned to the hodoscope.





\section{Summary}
\label{sec:summary}

A new electromagnetic calorimeter, BBCal, was constructed, installed, and operated successfully for the SBS program of high-$Q^2$ neutron form factor measurements in experimental Hall A at Jefferson Lab. The calorimeter met its design goals in terms of energy, position, and timing resolution. Overall, BBCal performed reliably and was invaluable for the success of the SBS physics program.

\section{Acknowledgments}
\label{sec:thanks}
We acknowledge the SBS collaboration who helped support the BBCal team and the JLab Hall A technical staff and Accelerator Operations group for providing outstanding support throughout the duration of the detector operation. We also like to specifically thank David Armstrong, Vladimir Goryachev, Mahlon Long, and Will Tireman for their work. This material is based upon work supported by the U.S. Department of Energy, Office of Science, Office of Nuclear Physics under contract DE-AC05-06OR23177 under which the Southeastern Universities Research Association (SURA) operates the Thomas Jefferson National Accelerator Facility for the United States Department of Energy. This work was also supported in part by the National Science Foundation contract PHY-2412825, the U.S. Department of Energy Office of Science, award ID DE-SC0021200, and Grant 21AG-1C085 by the Science Committee of the Republic of Armenia. 

\bibliographystyle{elsarticle-num}
\bibliography{references}

\end{document}